\documentclass[manuscript]{aastex631}

\usepackage{txfonts}
\usepackage{latexsym,bm}
\usepackage{url}
\usepackage{color}
\usepackage{colortbl}
\usepackage{multirow}
\newcommand{\dee}{\mathrm{d}}
\definecolor{mygray}{gray}{.9}

\usepackage[titletoc,title]{appendix}

\newcommand{\qinemail}{qingang@hit.edu.cn}
\newcommand{\hit}{School of Science, Harbin Institute of
Technology, Shenzhen, 518055, China}
\newcommand{\hitqin}{\hit; \qinemail}
\newcommand{\szlab}{Shenzhen Key Laboratory of Numerical Prediction for Space Storm,
Harbin Institute of Technology, 
Shenzhen, 518055, China}
\shorttitle{STATISTICAL QUANTITIES FOR SUBDIFFUSION}
\shortauthors{WANG AND QIN}

\begin{document}
\arraycolsep 0pt
	
\title{Relationship of
transport coefficients with statistical
quantities of 
charged particles}

\correspondingauthor{G. Qin}
\email{\qinemail}
	
\author[0000-0002-9586-093X]{J. F. Wang}
\affiliation{\hitqin}
	
\author[0000-0002-3437-3716]{G. Qin}
\affiliation{\hitqin}
\affiliation{\szlab}

\begin{abstract}

In previous studies, a general spatial transport
equation was derived from the Fokker-Planck equation. The latter
equation contains an infinite number of spatial derivative terms
$T_n=\kappa_{nz}\partial^n{F}/
\partial{z^n}$ 
with $n=1, 2, 3, \cdots$. 
Due to the complexity of
the general equation, 
some simplified equations 
with finitely many 
spatial derivative 
terms
have been used by 
astrophysical researches,
e.g., the diffusion equation, 
the hyperdiffusion equation, 
subdiffusion transport 
equation, etc.
In this paper, the simplified 
transport equations with the
spatial derivative 
terms up to the first-, second-, 
third-, fourth-, and fifth-order 
are listed,  
and their transport coefficient formulas 
are derived, respectively. 
We find that 
most of the transport 
coefficients
are determined by the 
corresponding 
statistical quantities.
In addition, 
we find that the well-known
statistical quantities, 
the skewness 
$\mathcal{S}$ and the kurtosis
$\mathcal{K}$, 
are determined 
by some transport coefficients.
The results can help 
one to use various transport
coefficients determined by
the statistical quantities, 
including 
many that are relatively new 
found in this paper, 
to study charged particle
parallel transport processes.

\end{abstract}


\section{INTRODUCTION}
\label{INTRODUCTION}

One of the most important 
problems in astrophysics and
space science
is the transport
of charged particles through
the magnetized plasmas in 
interplanetary and interstellar space
\citep{Jokipii1966, 
Schlickeiser2002, MatthaeusEA2003,
Shalchi2009, Shalchi2010, MalkovSagdeev2015, 
Shalchi2017, Shalchi2019, Shalchi2020a,
Shalchi2020b, Shalchi2021}. 
Due to the influence of turbulent magnetic field, 
charged particles perform a 
complex motion which can be seen as
the superposition of a deterministic
helical motion around the backgound
magnetic field lines and an irregular
component. 
Therefore, statistical description 
is necessary for describing charged particle motion 
and the methods of stochastic process
have been widely used
\citep{Schlickeiser2002, MatthaeusEA2003, Shalchi2009,
Shalchi2010, Shalchi2020b}.
In the past decades, 
the Fokker-Planck equation, which 
describes
the time evolution 
of the distribution function of
Brownian particles, is employed as the 
basis of the investigation of 
charged particle transport
\citep{Schlickeiser2002, Shalchi2005, 
Shalchi2009, LasuikEA2019,
ShalchiGammon2019, Shalchi2021b}. 

In previous investigations,
scientists have found that 
the spatial transport 
coefficients of 
charged energetic particles,
including the parallel and 
perpendicular
diffusion coefficients,
are the 
key parameters describing 
the modulation of galactic cosmic 
rays
\citep{Parker1965, BurgerEA1998, 
q2007,
Moraa2013, OughtonEngelbrecht2021, Potgieter2013, 
qz2014, qs2017, qw2018}, 
transport of solar energetic particles
\citep{ReamesEA1996, ReamesEA1997, 
Droege2000, ZankEA2000, QinEA2006,
QinEA2013}, 
diffusive acceleration of
charged particles by shocks
\citep{ZankEA2000, LiEA2003,
LiEA2005, ZankEA2006, DoschEA2010,
LiEA2012, HuEA2017}, etc.
Thus, the spatial 
transport coefficient formulas 
and corresponding 
spatial transport
equations
have to be obtained
\citep{Schlickeiser2002, SchlickeiserEA2008, 
Shalchi2009, wq2018, wq2019, 
Shalchi2021b}.

In the past few decades,
transport equations 
with finite spatial derivative terms,
e.g., the diffusion equation, 
the convection-diffusion 
equation and 
so on,
have drawn researchers' attentions
and been widely studied
\citep{Shalchi2009}.
In addition, 
in order to explore 
particle transport in turbulent 
plasmas, 
perturbation theory was 
employed by
\citet{MalkovSagdeev2015}
to 
obtain
the hyperdiffusion equation 
from the Fokker-Planck 
description. 
Meanwhile, 
\citet{ShalchiArendt2020} 
obtained a transport equation
with the fourth-order spatial 
derivative term for the subdiffusion process.
By integrating the Fokker-Planck 
equation over pitch angle,
\citet{wq2019} derived 
the general spatial
transport equation,
which contains an infinite
number of spatial derivative terms
$T_n=\kappa_{nz}\partial^n{F}/
\partial{z^n}$ 
with $n=1, 2, 3, \cdots$.  
Although the general spatial 
transport equation is a 
generalized
form of the transport equation
with finite spatial derivative terms 
and 
can describe more 
propagation processes,
this equation is too 
complex to be used in relevant
studies. 
In order to
describe various 
charged particle transport processes,
more simplified types 
of the general spatial transport
equations (STGEs)
should be 
thoroughly explored. 
Obviously, the equations of
hyperdiffusion and 
subdiffusion
are special cases
of the general transport 
equation and belong to 
the STGEs.  

Due to the preferred direction
induced by the background magnetic 
field,
parallel
transport of charged particles
is different
from cross-field one. 
For a variety of reasons, 
parallel transport is, 
in general, 
stronger than
perpendicular propagation. 
In the past,
a lot of progress has 
been achieved
in the analytical description of 
perpendicular diffusion
\citep{MatthaeusEA2003, Shalchi2010,
qs2014, Shalchi2017, Shalchi2019,
Shalchi2021},
but parallel transport 
is relatively 
poorly understood
\citep{Shalchi2022}. 
The problem
should be revisited and 
more investigations need to be
conducted. 
The topic
of this paper is to explore
the parallel transport coefficients
of various STGEs which describe
the 
parallel propagation of 
charged particles. 
For the parallel
diffusion coefficient, generally speaking, 
there are three different definitions
\citep{wq2019}, 
i.e., 
the displacement variance definition
$\kappa_{zz}^{DV}=\lim_{t\rightarrow t_{\infty}}d\sigma^2/(2dt)$
with the first- and second-order moments of charged particle 
distribution function,
the Fick's law definition
$\kappa_{zz}^{FL}=J/X$ with $X=\partial{F}/\partial{z}$,
and the TGK formula definition $\kappa_{zz}^{TGK}=\int_0^{\infty}dt
\langle v_z(t)v_z(0) \rangle$.
However, 
it has been demonstrated that 
for some scenarios 
different definitions of 
parallel diffusion 
are not equivalent.
\citet{wq2019} has proved that  
$\kappa_{zz}^{DV}$, rather than 
$\kappa_{zz}^{FL}$
and 
$\kappa_{zz}^{TGK}$, 
is the most appropriate definition. 
In this paper, the transport
coefficients expressed by the moments
of distribution function 
are derived, and their 
relations with statistical quantities
are explored.    

For convenience, 
if the highest order spatial
derivative term is $m$th-order, 
we call this simplified equation  
as the $m$th-order one.  
In this paper, we list 
all the 
simplified equations belonging to
the first-, second-, third-,
fourth-, and fifth-order 
spatial transport equations. 
The transport coefficients  
expressed by the moments of
distribution function 
can be obtained using 
the method proposed by
\citet{wq2019}. 
This paper is organized as follows. 
In Section 
\ref{The governing equations 
and transport regimes of 
the charged energetic particles}, 
the derivation of the
general spatial transport equation
is introduced, and 
the transport regimes are listed.  
In Sections \ref{The transport coefficient and statistical 
quantity of the first-order 
transport equation}, 
\ref{The transport coefficients 
and statistical 
quantities of the second-order 
transport equation},
\ref{The transport coefficients 
and statistical 
quantities of the third-order 
transport equation},
\ref{The transport coefficients 
and statistical 
quantities of the fourth-order 
transport equation},
and
\ref{The transport coefficients 
and statistical 
quantities of the fifth-order 
transport equation}, 
the transport coefficients 
and the corresponding statistical 
quantities of the first-,
second-, thrid-, fourth-, and
fifth-order spatial transport 
equations are derived, respectively.  
Meanwhile, 
the relationship
of transport coefficients with 
statistical quantities
are investigated, and 
the meanings of some common
statistical 
quantites are discussed. 
We conclude
and summarize our results in Section 
\ref{SUMMARY AND CONCLUSION}.

\section{The governing equations and transport regimes 
of the charged energetic particles}
\label{The governing equations and transport regimes 
	of the charged energetic particles}

\subsection{The general spatial
transport equation 
of charged particles}
\label{The general spatial
transport equation 
of charged particles}

The charged energetic particle 
transport
in the interplanetary and intersteller
plasmas is described by the well-known
Fokker-Planck equation,
which considers  
1-D spatial coordinate 
$z$
\citep{Schlickeiser2002, SchlickeiserEA2007, SchlickeiserEA2008,
Shalchi2009} 
\begin{equation}
\frac{\partial{f}}{\partial{t}}
+ v\mu\frac{\partial{f}}
{\partial{z}}=
\frac{\partial{}}{\partial{\mu}}
\left[D_{\mu \mu}(\mu)
\frac{\partial{f}}
{\partial{\mu}}-\frac{v}{2L}
(1-\mu^2)f \right],
\label{modified Fokker-Planck equation}
\end{equation}
Here, $f=f(z, \mu, t)$ is the distribution 
function of charged particles,
$t$ is time, $z$ is the spatial
coordiate, $\mu$ is pitch-angle cosine,
$v$ is the particle speed,
$L$ is the characteristic length
of adiabatic focusing effect,
and $D_{\mu\mu}(\mu)$ 
is the pitch-angle
diffusion coefficient as the function
of pitch-angle cosine $\mu$.  

If the gyrotropic
cosmic-ray 
density $f(z, \mu, t)$ in 
the phase space 
adjusts very quickly to
the quasi-equilibrium state 
through the pitch-angle diffusion,
the
distribution function $f(z, \mu, t)$ 
can be written as
the isotropic part $F(z,t)$ and the
anisotropic one $g(z,\mu,t)$
\citep
{SchlickeiserEA2007,
SchlickeiserEA2008, 
wq2018, wq2019}
\begin{equation}
f(z, \mu, t)=F(z, t)+g(z, \mu, t)
\label{f=F+g}
\end{equation}
with the normalization condition
\begin{equation}
F(z, t)=\frac{1}{2}\int_{-1}^1 d\mu f(z, \mu, t)
\label{F and f}
\end{equation}
and
\begin{equation}
\int_{-1}^1 d\mu g(z, \mu, t)=0.
\label{integrate g over mu=0}
\end{equation}
Here, the anisotropic part of
the distribution function
$f(z, \mu, t)$
can be found 
by the method developed by
\citet{HeEA2014} and \citet{wq2018, wq2019}
\begin{equation}
g(z,\mu,t)=L\left(\frac{\partial{F}}
{\partial{z}}
-\frac{F}{L}\right)\left[1-
\frac{2e^{M(\mu)}}{\int_{-1}^{1}
	d\mu e^{M(\mu) }}\right]
+e^{M(\mu)}\left[R(\mu,t)
-\frac{\int_{-1}^{1}d\mu
	e^{M(\mu)}R(\mu,t)}
{\int_{-1}^{1}d\mu
	e^{M(\mu) }}\right]
\label{g}
\end{equation}
with
\begin{eqnarray}
R(\mu,t)&=&\int_{-1}^{\mu} d\nu
e^{-M(\nu)}\Phi(\nu,t).
\label{R(mu)}
\end{eqnarray}
Integrating Equation
(\ref{modified Fokker-Planck equation})
over $\mu$ and using Equations
(\ref{F and f})-(\ref{g}),  
we can obtain 
the charged particle
transport equation in real space
\begin{eqnarray}
\frac{\partial{F}}{\partial{t}}=
&&\left(-\kappa_z'\frac{\partial{F}}
{\partial{z}}+\kappa_{zz}'
\frac{\partial^2{F}}{\partial{z^2}}
+\kappa_{3z}'
\frac{\partial^3{F}}{\partial{z^3}}
+\kappa_{4z}'
\frac{\partial^4{F}}{\partial{z^4}}
+\cdots\right)
+\left( \kappa_{tz}'
\frac{\partial^2{F}}{\partial{t}
	\partial{z}}
+ \kappa_{ttz}'\frac{\partial^3{F}}
{\partial{t^2}\partial{z}}
+ \kappa_{tttz}'\frac{\partial^4{F}}
{\partial{t^3}\partial{z}}
+\cdots\right)\nonumber\\
&&+\left( \kappa_{tzz}'
\frac{\partial^3{F}}{\partial{t}
	\partial{z^2}}
+ \kappa_{ttzz}'\frac{\partial^4{F}}
{\partial{t^2}\partial{z^2}}
+ \kappa_{tttzz}'\frac{\partial^5{F}}
{\partial{t^3}\partial{z^2}}
+\cdots\right)
+\cdots.
\label{equation of F with constant coefficient}
\end{eqnarray}
Here, $F=F(z,t)$ is the isotropic part 
of the distribution function $f(z,\mu,t)$
which satisfies the normalization condition (\ref{F and f}), and 
the parameters
$\kappa_z'$, $\kappa_{zz}'$,
$\kappa_{3z}'$, $\cdots$, 
$\kappa_{tzz}'$, $\cdots$
are the transport coefficients
corresponding to
different transport terms
in the latter transport equation.
With iterative operation on 
Equation 
(\ref{equation of F 
with constant coefficient}),
the terms containing time and space 
cross derivatives on the right-hand 
side
can be eliminated 
and the equation becomes
\begin{eqnarray}
\frac{\partial{F}}{\partial{t}}
&&=\sum_{n=1}^{\infty}
\kappa_{nz}\frac{\partial^n{F}}
{\partial{z^n}}
\label{governing equation in 
real space}
\\
&&
=\sum_{n=1}^{\infty}T_n
\end{eqnarray}
with 
$\kappa_{1z}=-\kappa_z$,
$\kappa_{2z}=\kappa_{zz}$,
$\kappa_{3z}=\kappa_{zzz}$,
$\cdots$, and 
\begin{eqnarray}
T_n=\kappa_{nz}\frac{\partial^n{F}}
{\partial{z^n}}.
\label{Tn}
\end{eqnarray}
Here, $\kappa_{nz}$ is
the transport coefficient
corresponding to the spatial
derivative term. 
Obviously, 
the first term on 
the right-hand side 
of Equation (\ref{governing equation in 
real space})
is the convection one, 
the second one describes 
the diffusion process. 
In the following subsection,
the effects 
of the other terms on the 
right-hand side
are investigated. 
Equation 
(\ref{governing equation in 
real space})
is the most general spatial 
transport description derived 
directly
from the Fokker-Planck equation, 
and forms the starting point 
of the research in this paper.

\subsection{The transport regimes of 
charged particles}
\label{The transport behavior of charged energetic particles}

For $\Delta z=z-z_0=z$
with $z_0=0$, 
the second- order 
moments of charged
particle distribution function 
$F(z,t)$ is shown as 
\begin{eqnarray}
\left\langle z^2 \right\rangle
=\int_{-\infty}^{+\infty}
\dee z z^2 F(z,t).
\end{eqnarray} 
According to the temporal 
behavior of the mean square 
displacement
\begin{eqnarray}
\left\langle z^2 \right\rangle
\sim t^{\sigma}, 
\end{eqnarray} 
the particle 
transport is characterized 
by different regimes \citep[e.g.,][]{Shalchi2009}
\begin{displaymath}
\left\{\begin{array}{ll}
	0<\sigma<1: & \textrm{ subdiffusion}\\
	\sigma=1: & \textrm{ Markovian 
		diffusion}\\
	1<\sigma<2: & \textrm{ superdiffusion}\\
	\sigma=2: & \textrm{ ballistic motion.}
	\end{array} \right.
\label{transport regime definitions}
\end{displaymath}
In the following, 
according to the latter definitions,
we explore the 
transport regimes 
described by the spatial derivative 
terms $T_n$ with $n=1, 2, 3, \cdots$
in Equation 
(\ref{governing equation in 
real space}).

\subsection{The transport regimes
described by $T_n$}
\label{The transport regime
described by Tn}

\subsubsection{The transport regime
described by $T_1$}
\label{The transport regime
described by T1}

With Equation (\ref{Tn}),
the formulas for
time derivative of
the first- and second-order 
moments of distribution function
caused by the convection term 
$T_1$ can be derived as  
\begin{eqnarray}
&&\frac{\dee}{\dee t}\left\langle 
z\right\rangle
=\kappa_{1z}\int_{-\infty}
^{+\infty}\dee z z 
\frac{\partial{F}}
{\partial{z}}
=\int_{-\infty}
^{+\infty}\dee z zT_1,\\
&&\frac{\dee}{\dee t}\left\langle 
z^2\right\rangle 
=\kappa_{1z}\int_{-\infty}
^{+\infty}\dee z z^2 
\frac{\partial{F}}{\partial{z}}
=\int_{-\infty}
^{+\infty}\dee z z^2T_1.
\end{eqnarray} 
Using partial integration
yields
\begin{eqnarray}
&&\frac{\dee}{\dee t}\left\langle 
z\right\rangle
=-\kappa_{1z},
\label{231-1}
\\
&&\frac{\dee}{\dee t}\left\langle 
z^2\right\rangle
=-2\kappa_{1z}
\left\langle 
z\right\rangle
\label{231-2}
\end{eqnarray} 
with the following 
normalization condition
\begin{eqnarray}
\int_{-\infty}^{+\infty}
dz F(z, t)=1.
\end{eqnarray} 
To proceed, from Equations
(\ref{231-1}) and (\ref{231-2}),
the first- and 
second-order moments produced
by the convection term $T_1$ are
shown as
\begin{eqnarray}
&&\left\langle 
z\right\rangle =-\kappa_{1z} t,\\
&&\left\langle 
z^2\right\rangle=-2\kappa_{1z}^2 t^2
\sim t^2,
\end{eqnarray} 
The latter formulas show that 
term $T_1$ describes
the ballistic regime.
 
\subsubsection{The transport regime
described by $T_2$}
\label{The transport regime
described by T2}

As in the previous section, 
the formulas for
time derivative of
the second-order
moment caused by the diffusion term 
is rewritten as follows
\begin{eqnarray}
\frac{\dee}{\dee t}\left\langle 
z^2\right\rangle
=\kappa_{2z}\int_{-\infty}
^{+\infty}\dee z z^2 
\frac{\partial^2{F}}{\partial{z^2}}
=\int_{-\infty}
^{+\infty}\dee z z^2 T_2.
\end{eqnarray}
Partial integration of the latter 
equation leads to 
the following expression
\begin{eqnarray}
\frac{\dee}{\dee t}\left\langle 
z^2\right\rangle=
\kappa_{2z}.
\label{232d/dt}
\end{eqnarray}  
Employing integrating 
Equation (\ref{232d/dt})
over time with zero initial 
condition
leads to
\begin{eqnarray}
\left\langle z^2\right\rangle =
\kappa_{2z}t,
\end{eqnarray} 
which indicates the term $T_2$
describes the Markovian 
diffusion regime.
 
\subsubsection{The transport regime
described by $T_n$ with $n=3,4,5,\cdots$}
\label{The transport regime
described by T3}

By using integration by parts
and considering formula
(\ref{Tn}),   
we can obtain 
the formulas for time derivative
of
the second-order
moments caused by the  
spatial derivative term $T_n$
with $n=3,4,5,\cdots$
\begin{eqnarray}
\frac{\dee}{\dee t}\left\langle 
z^2\right\rangle
=\kappa_{nz}\int_{-\infty}
^{+\infty}\dee z z^2 
\frac{\partial^n{F}}{\partial{z^n}}
=\int_{-\infty}
^{+\infty}\dee z z^2 T_n=0.
\label{<z^2> for T3}
\end{eqnarray}
According to the definitions 
shown in Subsection
\ref{The transport behavior of charged energetic particles},
the time derivative of 
the subdiffusion regime satisfies
\begin{eqnarray}
\frac{\dee}{\dee t}\left\langle 
z^2\right\rangle=t^{\sigma-1}
\end{eqnarray} 
For large enough time $t_a$ and 
with the subdiffusion regime 
$0<\sigma<1$, the latter formula becomes
\begin{eqnarray}
\frac{\dee}{\dee t}\left\langle 
z^2\right\rangle
=\lim_{t\to t_a}t^{\sigma-1}=0
\end{eqnarray} 
Comparing the latter equation with
Equation (\ref{<z^2> for T3}) we 
can find that 
the terms $T_n$ with 
$n=3,4,\cdots$ 
describe
subdiffusion process.  \\

Since the terms 
$T_1$, $T_2$, and  
$T_n$ with $=3,4,5\cdots$ 
belong to ballistic,
Markovian diffusion,
and 
a subdiffusive process,
respectively, 
we can infer that   
the different 
spatial derivative terms
$T_n$ in Equation 
(\ref{governing equation in 
real space}) 
correspond to different charged
particle 
transport regimes.

\subsection{The simplified types
of the general spatial transport equation}
\label{The simplified types
of the general spatial transport equation}

The general spatial transport 
equation 
(\ref{governing equation in 
real space}) 
derived rigorously from 
the Fokker-Planck equation
is highly complex and 
difficult to utilize.
In the past few decades, 
some simplified forms 
of the STGEs have  
been commonly used 
in astrophysical and laboratory 
plasma research. 
In order to describe various 
charged particle transport processes,
more simplified types 
of the STGEs should be 
thoroughly explored. 
Here, the following are 
some examples:

\begin{enumerate}
\item The convection equation
\begin{eqnarray}
	\frac{\partial{F}}{\partial{t}}
	=\kappa_{1z}\frac{\partial{F}}	
	{\partial{z}}=T_1
	\label{the convection equation-0}
\end{eqnarray}
which has the convection term $T_1$.
\item The diffusion equation
\begin{eqnarray}
\frac{\partial{F}}{\partial{t}}
=\kappa_{2z}
\frac{\partial^2{F}}{\partial{z^2}}
=T_2
\label{the diffusion equation-0}
\end{eqnarray}
which contains only the diffusion 
term $T_2$.
\item 
The hyperdiffusion equation
derived by \citet{MalkovSagdeev2015}
\begin{eqnarray}
\frac{\partial{F}}{\partial{t}}=
\kappa_{2z}\frac{\partial^2{F}}{\partial{z^2}}
+\kappa_{4z}\frac{\partial^4{F}}
{\partial{z^4}}=T_2+T_4,
\end{eqnarray}
which contains the spatial 
derivative 
terms $T_2$ and $T_4$. 
\item The spatial one-dimensional
subdiffusion equation
deduced by 
\citet{ShalchiArendt2020} 
for large enough 
time $t$ and $z$
\begin{eqnarray}
\frac{\partial{F}}
{\partial{t}}=\kappa_{4z}
\frac{\partial^4{F}}
{\partial{z^4}}=T_4
\end{eqnarray}
which contains only the term $T_4$. 
\end{enumerate}

In order to 
derive thoroughly     
the transport coefficients of 
the various STGEs,  
we classify the STGEs into 
different categories.  
If the highest order spatial
derivative term in an STGE
is the $m$th-order, 
the STGE is referred to 
as the $m$th-order spatial 
transport equation.   
Thus, all of the $m$th-order 
STGEs can be written as
\begin{eqnarray}
\frac{\partial{F}}{\partial{t}}=
\sum_{n=1}^{m}A_nT_n.
\label{transport equation in 
general form}
\end{eqnarray}
Here, 
\begin{equation}
	\left\{\begin{array}{ll}
		A_n=1 \textrm{ or }
		0 & \hspace{0.5cm} \textrm{ if } n<m\\
		A_n=1 & \hspace{0.5cm}\textrm{ if } n=m. 
	\end{array} \right.
	\label{Ai}
\end{equation}

The above discussion
can be summarised as: 
\begin{enumerate}
\item
The $m$th-order
spatial transport equation has
$2^{m-1}$ different STGEs. 
\item
The first-order
spatial transport equation 
has only one STGE,
i.e., the convection 
equation 
(see Section (\ref{The transport coefficient 
and statistical 
quantity of the first-order 
transport equation})). 
\item
The second-order
spatial transport equation 
contains two different STGEs,
i.e., the diffusion equation
and the 
convection-diffusion equation
(see Section (\ref{The transport 
coefficients and statistical 
quantities of the second-order 
transport equation})). 
\item
The third-order
spatial transport equation 
has
four different STGEs with derivation in
Section
\ref{The transport coefficients 
and statistical 
quantities of the third-order 
transport equation} and results 
in Table 
\ref{The transport coefficients and 
statistical quantities of the 
third-order transport equation-Table}. 
\item
The fourth-order
spatial transport equation 
has eight different 
STGEs,
with the derivation shown in
Section
\ref{The transport coefficients 
and statistical 
quantities of the fourth-order 
transport equation} and results in Table
\ref{The transport coefficients and statistical quantities of the 
fourth-order transport equation-Table}
\item
The fifth-order
spatial transport equation 
has sixteen different
STGEs,
with the results shown
in Table 
\ref{The transport 
coefficients  and 
statistical quantities of the 
fifth-order transport equation-table}. 
\item
For the $m$th order 
spatial transport equation with
$m\ge 6$, the derivation
is 
too complicated and, 
therefore, we do not
evaluate the transport coefficients
in this paper.
\end{enumerate}

\subsection{The convection and diffusion coefficients with the corresponding 
	statistical quantites}
\label{The convection and diffusion coefficients and corresponding 
	statistical quantites}

The well-known 
convection-diffusion equation,
which contains the convection and
diffusion terms, is as follows
\begin{eqnarray}
	\frac{\partial{F}}{\partial{t}}
	=\kappa_{1z}\frac{\partial{F}}
	{\partial{z}}
	+\kappa_{2z}
	\frac{\partial^2{F}}{\partial{z^2}}.
	\label{the diffusion-convection equation-0}
\end{eqnarray}
Here, $\kappa_{1z}$
is the convection coefficient
and $\kappa_{2z}$ is the diffusion 
coefficient.
From Equation 
(\ref{the diffusion-convection equation-0}), 
the formulas for time derivative
of
the first- and second-order moments
of the charged particle
distribution function
can be obtained
\begin{eqnarray}
	&&\frac{\dee}{\dee t}\langle z\rangle
	=\int_{-\infty}^\infty dz z\frac{\partial{F}}
	{\partial{t}}
	=\int_{-\infty}^\infty dzz
	\left(\kappa_{1z}\frac{\partial{F}}
	{\partial{z}}+\kappa_{2z}
	\frac{\partial^2{F}}{\partial{z^2}}\right)
	=-\kappa_{1z},\\
	&&\frac{\dee}{\dee 
		t}\left\langle z^2\right\rangle
	=\int_{-\infty}^\infty dz z^2\frac{\partial{F}}
	{\partial{t}}
	=\int_{-\infty}^\infty dzz^2
	\left(\kappa_{1z}\frac{\partial{F}}
	{\partial{z}}+\kappa_{2z}
	\frac{\partial^2{F}}{\partial{z^2}}\right)
	=
	-2\kappa_{1z}\langle z\rangle
	+2\kappa_{2z}.
	\label{ddt Delta z2 for diffusion-convection equation}
\end{eqnarray}
Combining the latter equations, 
one can find
\begin{eqnarray}
	&&\kappa_{1z}=\frac{\dee}{\dee t}
	\alpha^1_{11},
	\label{ddt Delta z for the streaming equation-0} \\
	&&\kappa_{2z}=
	\frac{1}{2}\frac{\dee}{\dee t}
	\alpha^2_{11}
	\label{kzz definied by 
		mean square displacement-0}
\end{eqnarray}
with 
\begin{eqnarray}
&&\alpha^1_{11}
=-\langle z\rangle, \\
&&\alpha^2_{11}=\left\langle z^2\right\rangle
-\langle z\rangle^2.
\label{beta12}
\end{eqnarray} 
Here, 
$\langle z\rangle$ is the mathematical expectation
of charged 
particle
distribution function, 
and $\left\langle z^2\right\rangle
-\langle z\rangle^2=
\left\langle \left(z-
\langle z\rangle\right)^2
\right\rangle$
is the second-order central
moment 
of charged 
particle
distribution function 
which is also called as
variance.
Obviously, 
the statistical quantities
$\alpha^1_{11}$ is 
negative mathematical
expectation,
and 
$\alpha^2_{11}$
is variance.
Equations 
(\ref{ddt Delta z for the streaming equation-0})
and
(\ref{kzz definied by 
	mean square displacement-0})
show that
the convection and diffusion coefficients
are determined by 
the statistical quantities 
$\alpha^1_{11}$ and $\alpha^2_{11}$,
respectively. 
For convenience, we use the 
notations of the statistical
quantities as the follows:
\begin{enumerate}
	\item 
	We use $\alpha^n_{s}$ to indicate the
	statistical quantity
	determining $\kappa_{nz}$.
	\item 
	In order to distinguish 
	the same transport coefficients
	in different transport 
	equation, 
	we set the subscript of  $\alpha^n_{s}$, i.e., $s$,
	as $A_1A_2\cdots A_i\cdots A_m$.
	Here, $A_i$ is defined in 
	Equation 
	(\ref{Ai}). 
\end{enumerate}

In Subsection \ref{The 
convection and diffusion 
coefficients and corresponding 
statistical quantites},
it is demonstrated 
{\bf}
that 
tranport  coefficients $\kappa_{1z}$
and $\kappa_{2z}$
are determined by mathematical
expectation and variance, respectively.
In the following sections,
we derive 
the transport coefficient formulas 
for different STGEs 
and explore the relationship of 
transport coefficients
with statistical quantities.

\section{The transport coefficient 
and statistical 
quantity of the first-order transport 
equation}
\label{The transport coefficient 
and statistical 
quantity of the first-order transport 
equation}

As summarized in Subsection
\ref{The simplified types
of the general spatial transport 
equation}, 
the first-order
transport equation has only one 
STGE, i.e., 
the convection equation
\begin{eqnarray}
\frac{\partial{F}}{\partial{t}}=
\kappa_{1z}\frac{\partial{F}}
{\partial{z}},
\label{streaming equation}
\end{eqnarray}
which describes the convection
transport process of charged particles
with a constant speed.  
Using the same approch as in
Subsection 
\ref{The convection and diffusion coefficients and corresponding 
statistical quantites},
the formula for the 
coefficient $\kappa_{1z}$
can be obtained as
\begin{eqnarray}
\kappa_{1z}=\frac{\dee}{\dee t}\alpha^1_1,
\label{ddt Delta z for the streaming equation}
\end{eqnarray}
with 
\begin{eqnarray}
\alpha_1^1=-\langle z\rangle.
\label{alpha1}
\end{eqnarray}
The latter two equations 
show that
the convection coefficient 
$\kappa_{1z}$ in the convection
equation 
is determined by the 
statistical quantity
$\alpha_1^1$, i.e.,
negative mathematical expectation.

\section{The transport coefficients 
and statistical 
quantities of the second-order 
transport equations}
\label{The transport coefficients 
and statistical 
quantities of the second-order 
transport equation}

The second-order transport equations
contains two different  
STGEs, i.e., the diffusion one
and convection-diffusion one.  

\subsection{The transport coefficient and the corresponding statistical
quantity of 
the diffusion equation}
\label{The statistical quantity for the diffusive equation}

The diffusion equation
\begin{eqnarray}
\frac{\partial{F}}{\partial{t}}=
\kappa_{2z}
\frac{\partial^2{F}}{\partial{z^2}} 
\label{the diffusion equation}
\end{eqnarray}
is one of the most well-known 
differential 
equations in the fields of 
mathematics
and physics.
The formulas for time 
derivative of
the first- and second-order
moments can be found easily
\begin{eqnarray}
&&\frac{\dee}{\dee t}\langle z\rangle
=0,
\label{ddt Delta z for diffusion equation} \\
&&\frac{\dee}{\dee t}
\left\langle z^2\right\rangle
=2\kappa_{2z}. 
\label{ddt Delta z2 for 
diffusion equation}
\end{eqnarray}
To proceed, the diffusion coefficient
can be written as
\begin{eqnarray}
\kappa_{2z}=\frac{1}{2}
\frac{\dee}{\dee t}
\alpha^2_{01}
\label{kzz formula 
in diffusion equation}
\end{eqnarray}
with
\begin{eqnarray}
\alpha^2_{01}=\left\langle z^2\right\rangle.
\label{beta02}
\end{eqnarray}
It is obvious that the diffusion
coefficient is determined by
the statistical quantity $\alpha^2_{01}$.

\subsection{The transport coefficients and the corresponding statistical
quantities of 
the convection-diffusion equation}
\label{The statistical quantity for the diffusive convective equation}

The second type of the general 
transport equation is 
the convection-diffusion
equation.
As shown in Subsection
\ref{The convection and diffusion coefficients and corresponding 
statistical quantites},
the convection coefficient 
$\kappa_{1z}$
and the diffusion one $\kappa_{2z}$
are determined by 
statistical quantities
mathematical expectation 
$\alpha^1_{11}$
and
variance $\alpha^2_{11}$, respectively. 
In addition, by comparing 
Equations 
(\ref{kzz definied by 
mean square displacement-0})
with 
(\ref{kzz formula 
in diffusion equation}),
we can find that 
the diffusion coefficient formulas
in the convection-diffusion equation
and the diffusion equation 
are different. \\

\section{The transport coefficients 
and statistical 
quantities of the third-order 
transport equations}
\label{The transport coefficients 
and statistical 
quantities of the third-order 
transport equation}

According to the discussion
in Subsection
\ref{The simplified types
of the general spatial transport 
equation}, 
for the third-order transport equations,
there are four different STGEs.
In this section, 
the formulas for the 
transport coefficient
of each STGE
are deduced and the 
corresponding 
statistical quantites
are investigated. 
The transport coefficients 
and corresponding statistical
quantities are listed in Table
\ref{The transport coefficients and 
	statistical quantities of the 
	third-order transport equation-Table}.

\subsection{The transport 
coefficient formulas and the corresponding statistical
quantities of the equation
with $T_1$, $T_2$, and $T_3$}
\label{The statistical quantity 
for the third-order differential equation}

Here, we investigate 
the transport equation 
\begin{eqnarray}
\frac{\partial{F}}{\partial{t}}
=
\kappa_{1z}\frac{\partial{F}}
{\partial{z}}
+\kappa_{2z}\frac{\partial^2{F}}
{\partial{z^2}}
+\kappa_{3z}\frac{\partial^3{F}}
{\partial{z^3}},
\label{equation123}
\end{eqnarray}
which contains the convection term
$T_1$, 
diffusion term $T_2$,
and subdiffusion term $T_3$. 
The formulas for time 
derivative of
the first-, second-, and 
third-order moments of 
the distribution function 
can easily be obtained 
via
\begin{eqnarray}
&&\frac{\dee}{\dee t}\langle z\rangle
=-\kappa_{1z},
\\
&&\frac{\dee}{\dee t}
\langle z^2\rangle
=
-2\kappa_{1z}\langle z\rangle
+2\kappa_{2z},
\\
&&\frac{\dee}{\dee t}
\langle z^3\rangle
=-3\kappa_{1z} \left\langle z^2 \right\rangle
+6\kappa_{2z}\left\langle z \right\rangle
-6\kappa_{3z}.
\end{eqnarray}
Combining the latter formulas
yields
\begin{eqnarray}
&&\kappa_{1z}=
\frac{\dee}{\dee t}
\alpha^1_{111},
\label{kz=alpha1}\\
&&\kappa_{2z}=
\frac{1}{2}\frac{\dee}{\dee t}
\alpha^2_{111},
\label{kzz=alpha2}\\
&&\kappa_{3z}=\frac{1}{6}
\frac{\dee}{\dee t}\alpha^3_{111}
\label{k3z=alpha3}
\end{eqnarray}
with
\begin{eqnarray}
&&\alpha^1_{111}=-\langle z\rangle.\\
&&\alpha^2_{111}=\left\langle z^2 \right\rangle
-\langle z\rangle^2,
\label{beta123}\\
&&\alpha^3_{111}=
3\left\langle z \right\rangle\left\langle z^2 \right\rangle
-2\left\langle z \right\rangle^3
-\left\langle z^3 \right\rangle.
\label{delta111}
\end{eqnarray}
Here, $\alpha^1_{111}$ 
and $\alpha^2_{111}$
are the negative mathematical
expectation and variance, 
respectively. 
If we rewrite
$\alpha^3_{111}$ as
\begin{eqnarray}
\alpha^3_{111}
=-\left\langle \left(z
-\langle z \rangle \right)^3 \right\rangle, 
\label{delta111 expressed 
by 3rd order
central moment}
\end{eqnarray}
we can find that $\alpha^3_{111}$
is the negative third-order 
central moment.  
Thus, Equations 
(\ref{kz=alpha1})-(\ref{delta111 
expressed by 3rd order
central moment})
demonstrate that
the transport coefficients
$\kappa_{1z}$, $\kappa_{2z}$,
and $\kappa_{3z}$ in 
Equation (\ref{equation123})
are determined by the 
corresponding statistical
quantities, i.e.,
negative mathematical 
expectation
$\alpha^1_{111}$, variance $\alpha^2_{111}$,
and the negative third-order 
central moment $\alpha^3_{111}$,
respectively. 

To continue, with Equations 
(\ref{beta123}) and 
(\ref{delta111}),
the skewness
of charged particle
distribution function,
which describes the 
uniformity of the distribution
function, 
can be obtained as 
\begin{eqnarray}
\mathcal{S}
=\frac{\left\langle \left(z
-\langle z \rangle \right)^3 \right\rangle}
{\left\langle \left(z
-\langle z \rangle  \right)^2\right\rangle^{3/2}}
=\frac{\alpha^3_{111}}
{{\alpha^2_{111}}^{3/2}}
=-\frac{\sqrt{2}}{3}
\frac{\kappa_{3z}}
{\kappa_{2z}^{3/2}}
\frac{1}{\sqrt{t}}.
\end{eqnarray}
The latter formula shows that
the skewness $\mathcal{S}$
is a function of 
the transport coefficients
$\kappa_{2z}$ and $\kappa_{3z}$.

\subsection{The transport coefficient formula and the corresponding statistical
quantity of the equation
with $T_3$}
\label{The statistical 
quantity for the equation 3-order}

In this subsection, 
we investigate 
the transport coefficient
of the transport equation
\begin{eqnarray}
\frac{\partial{F}}{\partial{t}}
=
\kappa_{3z}\frac{\partial^3{F}}
{\partial{z^3}},
\label{equation003}
\end{eqnarray}
which contains only one term
$T_3$ on the right-hand. With the method used in the 
paper of \citet{wq2019},
it is straightforward to obtain
\begin{eqnarray}
&&\frac{\dee}{\dee t}\langle z\rangle	=0, \\
&&\frac{\dee}{\dee t}\langle z^2\rangle
=0, \\
&&\frac{\dee}{\dee t}
\langle z^3\rangle
=-6\kappa_{3z}. 
\label{ddt Delta z3 for 003}
\end{eqnarray}
From Equation 
(\ref{ddt Delta z3 for 003})
we can find 
the subdiffusion transport 
coefficient $\kappa_{3z}$ as
\begin{eqnarray}
\kappa_{3z}=\frac{1}{6}
\frac{\dee}{\dee t}
\alpha^3_{001}
\label{k3z for equation003}
\end{eqnarray}
with
\begin{eqnarray}
\alpha^3_{001}=-\left\langle z^3\right\rangle.
\label{delta003}
\end{eqnarray}
Here, $\left\langle z^3\right\rangle$
is the third-order moment of distribution function of charged 
particles. 
Equations (\ref{k3z for equation003})
and 
(\ref{delta003})
denote that the transport 
coefficient $\kappa_{3z}$ 
can be expressed as a function 
of statistical quantity 
$\alpha^3_{001}$.

\subsection{The transport coefficient formulas and the corresponding statistical
quantities of the equation
with $T_1$ and $T_3$}
\label{The statistical quantity for the equation 1-order+3-order}

The third STGE of the spatial 
third-order 
tansport equation is shown as
\begin{eqnarray}
\frac{\partial{F}}{\partial{t}}=
\kappa_{1z}\frac{\partial{F}}
{\partial{z}}
+\kappa_{3z}\frac{\partial^3{F}}{\partial{z^3}},
\label{equation103}
\end{eqnarray}
which contains $T_1$ and $T_3$. 
As done in the previous subsections,
the formulas for time 
derivative of 
the moments
of distribution function can be 
derived as 
\begin{eqnarray}
&&\frac{\dee}{\dee t}
\left\langle z\right\rangle
=-\kappa_{1z},
\label{ddt Delta z in 101} 
\\
&&\frac{\dee}{\dee t}
\left\langle z^2\right\rangle
=
-2\kappa_{1z}\langle z\rangle,
\\
&&\frac{\dee}{\dee t}
\left\langle z^3\right\rangle
=
-3\kappa_{1z} \left\langle z^2 
\right\rangle
-6\kappa_{3z}. 
\label{ddt Delta z3 in 101}
\end{eqnarray}
To continue, the transport 
coefficient formulas
can be derived from Equations
(\ref{ddt Delta z in 101})
and
(\ref{ddt Delta z3 in 101}) as
\begin{eqnarray}
&&\kappa_{1z}=\frac{\dee}{\dee t}
\alpha^1_{101},
\label{kz103}\\
&&\kappa_{3z}=\frac{1}{6}
\frac{\dee}{\dee t}
\alpha^3_{101},
\label{k3z103}
\end{eqnarray}
with
\begin{eqnarray}
&&\alpha^1_{101}=
-\left\langle z\right\rangle,
\label{alpha103}\\
&&\alpha^3_{101}=3\langle z\rangle
\left\langle z^2 \right\rangle
-2\left\langle z 
\right\rangle^3
-\left\langle z^3 \right\rangle. 
\label{delta103}
\end{eqnarray}
It is obvious that the transport
coefficients $\kappa_{1z}$
and $\kappa_{3z}$ are expressed
in terms of statistical quantities
$\alpha^1_{101}$ and 
$\alpha^3_{101}$, respectively.

\subsection{The transport coefficient formulas and the corresponding statistical
quantities of the equation
with $T_2$ and $T_3$}
\label{The statistical quantity 
for the equation 2-order+3-order}

The last STGE of the spatial
third-order equation is shown as
\begin{eqnarray}
\frac{\partial{F}}{\partial{t}}
=
\kappa_{2z}\frac{\partial^2{F}}
{\partial{z^2}}
+\kappa_{3z}\frac{\partial^3{F}}
{\partial{z^3}},
\label{equation023}
\end{eqnarray}
which has $T_2$ and $T_3$. 
The formulas 
for the transport coefficients 
can be easily derive as 
\begin{eqnarray}
&&\kappa_{2z}=
\frac{1}{2}\frac{\dee}{\dee t}
\alpha^2_{011},
\label{kzz023} \\
&&
\kappa_{3z}=\frac{1}{6}	\frac{\dee}{\dee t}\alpha^3_{011}
\label{k3z023}
\end{eqnarray}
with
\begin{eqnarray}
&&\alpha^2_{011}=\left\langle z^2\right\rangle,\\
&&\alpha^3_{011}=
3\left\langle z \right\rangle
\left\langle z^2\right\rangle
-\left\langle z^3\right\rangle. 
\end{eqnarray}
It is demonstrated that 
the transport coefficients
$\kappa_{2z}$ 
and $\kappa_{3z}$
are expressed as the time 
derivative of 
the statistical quantities
$\alpha^2_{011}$ and 
$\alpha^3_{011}$, respectively.

\section{The transport coefficients 
and statistical 
quantities of the fourth-order 
transport equation}
\label{The transport coefficients 
and statistical 
quantities of the fourth-order 
transport equation}

As shown in Subsection
\ref{The simplified types
of the general spatial 
transport equation},
the fourth-order transport 
equation has 
eight different STGEs,
which include the hyperdiffusion
equation derived by \citet{MalkovSagdeev2015}
and the subdiffusion transport 
equation deduced by 
\citet{ShalchiArendt2020}. 
The relationships of 
transport coefficients 
with statistical
quantities are listed in Table
\ref{The transport coefficients and 
	statistical quantities of the 
	fourth-order transport equation-Table}.

\subsection{The transport coefficient formulas and the corresponding statistical
quantities of the equation 
with $T_1$, $T_2$, $T_3$, and 
$T_4$}
\label{The statistical quantity for the fourth-order differential equation}

The first fourth-order STGE
is shown as
\begin{eqnarray}
\frac{\partial{F}}{\partial{t}}=
\kappa_{1z}\frac{\partial{F}}
{\partial{z}}
+\kappa_{2z}\frac{\partial^2{F}}{\partial{z^2}}
+\kappa_{3z}\frac{\partial^3{F}}{\partial{z^3}}
+\kappa_{4z}\frac{\partial^4{F}}{\partial{z^4}},
\label{equation1234}
\end{eqnarray}
which contains the terms 
$T_1$, $T_2$, $T_3$ and 
$T_4$. 
The transport coefficients of
the latter equation can be 
found as 
\begin{eqnarray}
&&\kappa_{1z}=\frac{\dee}{\dee t}
\alpha^1_{1111},
\\
&&\kappa_{2z}=
\frac{1}{2}\frac{\dee}{\dee t}
\alpha^2_{1111},
\\
&&\kappa_{3z}=	\frac{1}{6}\frac{\dee}{\dee t}
\alpha^3_{1111},
\\
&&\kappa_{4z}=\frac{1}{24}\frac{\dee}
{\dee t}
\alpha^4_{1111}
\end{eqnarray}
with
\begin{eqnarray}
&&\alpha^1_{1111}=-\langle z\rangle,\\
&&\alpha^2_{1111}=\left\langle z^2\right\rangle
-\langle z\rangle^2 ,	\\
&&\alpha^3_{1111}=
3\left\langle z \right\rangle\left\langle z^2 \right\rangle
-\left\langle z^3\right\rangle
-2\left\langle z \right\rangle^3,\\
&&\alpha^4_{1111}=\left\langle z^4
\right\rangle
+12
\left\langle z^2\right\rangle\langle z\rangle^2
-4
\langle z\rangle\left\langle z^3\right\rangle
-3\left\langle z^2\right\rangle^2
-6\left\langle z \right\rangle^4. 
\end{eqnarray}	
Here, the statistical quantity
$\alpha^4_{1111}$ 
determines the subdiffusion
coefficient $\kappa_{4z}$.  
The statistical quantities
$\alpha^1_{1111}$, $\alpha^2_{1111}$,
and $\alpha^3_{1111}$ in Equation
(\ref{equation1234})
are equal to  
$\alpha^1_{111}$, $\alpha^2_{111}$,
and $\alpha^3_{111}$ in Equation
(\ref{equation123}), respectively. 
Accordingly, 
the transport coefficients
$\kappa_{1z}$, $\kappa_{2z}$,
$\kappa_{3z}$, and $\kappa_{4z}$
in Equation (\ref{equation1234})
are identical with the corresponding
coefficients in Equation
(\ref{equation123}). 
That is, the term $T_4$ 
in Equation
(\ref{equation1234})
has no 
influence on the formulas
for $\kappa_{1z}$, $\kappa_{2z}$,
$\kappa_{3z}$ and $\alpha^1_{1111}$, $\alpha^2_{1111}$,
and $\alpha^3_{1111}$. 
In addition, the transport coefficients
$\kappa_{1z}$, $\kappa_{2z}$,
$\kappa_{3z}$ and $\kappa_{4z}$ 
are determined by statistical 
quantities $\alpha^1_{1111}$, $\alpha^2_{1111}$, $\alpha^3_{1111}$,
and $\alpha^4_{1111}$, respectively. 

\subsection{The transport coefficient formulas and the corresponding statistical
quantities of the equation
with $T_2$, $T_3$, and $T_4$}

The second STGE is as follows
\begin{eqnarray}
\frac{\partial{F}}{\partial{t}}=
\kappa_{2z}\frac{\partial^2{F}}{\partial{z^2}}
+\kappa_{3z}\frac{\partial^3{F}}{\partial{z^3}}
+\kappa_{4z}\frac{\partial^4{F}}{\partial{z^4}},
\label{234}
\end{eqnarray}
which includes $T_2$, $T_3$, and
$T_4$. 
As done in the above 
subsections, the transport 
coefficients can be found as
\begin{eqnarray}
&&\kappa_{2z}=\frac{1}{2}
\frac{\dee}{\dee t}
\alpha^2_{0111},\\
&&\kappa_{3z}=
\frac{1}{6}\frac{\dee}{\dee t}
\alpha^3_{0111},\\
&&\kappa_{4z}=\frac{1}{24}
\frac{\dee}{\dee t}\alpha^4_{0111}.
\end{eqnarray}
with
\begin{eqnarray}
&&\alpha^2_{0111}=	
\left\langle z^2\right\rangle,\\
&&\alpha^3_{0111}=3\left\langle z \right\rangle
\left\langle z^2 \right\rangle 
-\left\langle z^3 \right\rangle,\\
&&\alpha^4_{0111}=
\left\langle z^4\right\rangle
-3\left\langle z^2\right\rangle^2
-4\langle z\rangle 
\left\langle z^3\right\rangle
+12\langle z\rangle^2
\left\langle z^2\right\rangle.
\end{eqnarray}	
Here, we can find 
$\alpha^2_{0111}=\alpha^2_{011}
=\alpha^2_{01}$, and $\alpha^3_{0111}
=\alpha^3_{011}$. What's more, 
the transport coefficients 
$\kappa_{2z}$, $\kappa_{3z}$,
and $\kappa_{4z}$ are determined 
by $\alpha^2_{0111}$, 
$\alpha^3_{0111}$,
and $\alpha^4_{0111}$, respectively.

\subsection{The transport coefficient formulas 
and the corresponding statistical
quantities 	
of the equation with $T_3$ and $T_4$}

The transport equation with 
two subdiffusion terms $T_3$
and $T_4$ is shown as 
\begin{eqnarray}
\frac{\partial{F}}{\partial{t}}=
\kappa_{3z}\frac{\partial^3{F}}{\partial{z^3}}
+\kappa_{4z}\frac{\partial^4{F}}{\partial{z^4}}.
\end{eqnarray}
The relations
of transport coefficients
and statistical quantities
can be obtained as
\begin{eqnarray}
&&\kappa_{3z}=
\frac{1}{6}\frac{\dee}{\dee t}
\alpha^3_{0011},\\
&&\kappa_{4z}=\frac{1}{24}
\frac{\dee}{\dee t}
\alpha^4_{0011}
\end{eqnarray}
and 
\begin{eqnarray}
&&\alpha^3_{0011}=
-\left\langle z^3\right\rangle,
\\
&&\alpha^4_{0011}
=\left\langle z^4\right\rangle
-4\langle z\rangle 
\left\langle z^3\right\rangle. 
\end{eqnarray}
Comparing with 
the results 
of the previous subsections,
we can find that $\alpha^4_{0011}$
is relatively new 
and $\alpha^3_{0011}=\alpha^3_{001}$.
It is obvious that the statistical 
quantities $\alpha^3_{0011}$
and $\alpha^4_{0011}$
determine $\kappa_{3z}$ 
and $\kappa_{4z}$, respectively.

\subsection{The transport coefficient formula and the corresponding statistical
quantity of the equation with $T_4$}

For the perpendicular
subdiffusion process
$\left\langle x^2\right\rangle
\sim \sqrt{t}$, through 
lengthy derivation 
\citet{ShalchiArendt2020} 
found its governing equation. 
In fact, the deducation is 
also applicable
to the parallel transport.
For 
$\left\langle z^2\right\rangle
\sim \sqrt{t}$ with
large enough time $t$ and $z$, 
the corresponding 
controlling equation is shown as
\begin{eqnarray}
\frac{\partial{F}}{\partial{t}}=
\kappa_{4z}\frac{\partial^4{F}}{\partial{z^4}},
\end{eqnarray}
which contains only one subdiffusion
term $T_4$. 
Accordingly,  
the transport coefficent
can be derived as 
\begin{eqnarray}
\kappa_{4z}=\frac{1}{24}
\frac{\dee}{\dee t}
\alpha^4_{0001}
\end{eqnarray}
with
\begin{eqnarray}
\alpha^4_{0001}
=\left\langle z^4\right\rangle. 
\end{eqnarray}
We can find that 
the statistical quantity
$\alpha^4_{0001}$ is relatively
new and it 
determines the subdiffusion
coefficient $\kappa_{4z}$.

\subsection{The transport coefficient formulas and the corresponding statistical
quantities of the equation
with $T_1$, $T_3$, and $T_4$}

In this part, the transport 
equation 
including the convection term $T_1$,
and the subdiffusion terms $T_3$ 
and $T_4$ is
explored, which is shown as 
\begin{eqnarray}
\frac{\partial{F}}{\partial{t}}=
\kappa_{1z}\frac{\partial{F}}
{\partial{z}}
+\kappa_{3z}\frac{\partial^3{F}}{\partial{z^3}}
+\kappa_{4z}\frac{\partial^4{F}}{\partial{z^4}}. 
\label{equation1034}
\end{eqnarray}
With the method used in the previous
subsections, we find 
\begin{eqnarray}
&&\kappa_{1z}=
\frac{\dee}{\dee t}\alpha^1_{1011},
\\
&&\kappa_{3z}=
\frac{1}{6}\frac{\dee}{\dee t}
\alpha^3_{1011}, \\
&&\kappa_{4z}
=\frac{1}{24}\frac{\dee}{\dee t}
\alpha^4_{1011}
\end{eqnarray}
with
\begin{eqnarray}
&&\alpha^1_{1011}=-\langle z\rangle,\\
&&\alpha^3_{1011}=3\langle z\rangle
\left\langle z^2 \right\rangle
-2\left\langle z 
\right\rangle^3
-\left\langle z^3 \right\rangle,\\
&&\alpha^4_{1011}=
\left\langle z^4 \right\rangle
-4\left\langle z^3\right\rangle\langle z\rangle
+6\left\langle z^2 \right\rangle\langle z\rangle^2
-3
\left\langle z \right\rangle^4.
\end{eqnarray}
Here, the statistical quantity
$\alpha^4_{1011}$
can be rewritten as
\begin{eqnarray}
\alpha^4_{1011}=
\left\langle \left(z
-\left\langle z \right\rangle
\right)^4 \right\rangle.
\end{eqnarray}
Here, the right-hand side of 
the latter equation 
is the fourth-order 
central
moment of charged particle
distribution function. 
It can be seen that the transport 
coefficients $\kappa_{1z}$,
$\kappa_{3z}$, and $\kappa_{4z}$
are determined by 
$\alpha^1_{1011}$, $\alpha^3_{1011}$,
and $\alpha^4_{1011}$, respectively.  
Additionly, we can find that
$\alpha^1_{1011}=\alpha^1_{101}$,
$\alpha^3_{1011}=\alpha^3_{101}$,
and $\alpha^4_{1011}=
\alpha^4_{101}$. That is, the higher order
term $T_4$ does not influence the
relathonships of the lower order
transport coefficients
with 
statistical quantities.

\subsection{The transport coefficient formulas and the corresponding statistical
quantities of the equation
with $T_1$, $T_2$, and $T_4$}

For the following transport equation with 
$T_1$, $T_2$, and $T_4$
\begin{eqnarray}
\frac{\partial{F}}{\partial{t}}=
\kappa_{1z}\frac{\partial{F}}
{\partial{z}}
+\kappa_{2z}\frac{\partial^2{F}}{\partial{z^2}}
+\kappa_{4z}\frac{\partial^4{F}}{\partial{z^4}},
\label{equation1204}
\end{eqnarray}
we can obtain easily the 
transport coefficients 
\begin{eqnarray}
&&\kappa_{1z}=
\frac{d}{dt}\alpha^1_{1101},\\
\label{ddt Delta z of equation1204}
&&\kappa_{2z}=
\frac{1}{2}\frac{\dee}{\dee t}
\alpha^2_{1101},\\
&&\kappa_{4z}
=\frac{1}{24}\frac{\dee}{\dee t}
\alpha^4_{1101}
\end{eqnarray}
with
\begin{eqnarray}
&&\alpha^1_{1101}=-\langle z\rangle,
\label{alpha1204}  \\
&&\alpha^2_{1101}=\left\langle z^2\right\rangle
-\langle z\rangle^2 ,	
\label{beta1204}  \\
&&\alpha^4_{1101}=
\langle z^4\rangle
-4\langle z\rangle\left\langle z^3\right\rangle
+12\left\langle z^2 \right\rangle\langle z\rangle^2
-3\langle z^2\rangle^2
-6\langle z\rangle^4. 
\label{delta1204}
\end{eqnarray}
Here, the statistical quantity 
$\alpha^4_{1101}$ is relatively 
new. In addition, 
$\alpha^1_{1101}=\alpha^1_{11}$
and $\alpha^2_{1101}=\alpha^2_{11}$
hold. 
In addition, 
the following formula can be
found
\begin{eqnarray}
\alpha^4_{1101}+3\left(\alpha^1_{1101}\right)^2
=\left\langle \left(z-\langle z \rangle \right)^4 \right\rangle,
\label{fourth order central moment
with alpha1101}
\end{eqnarray}
which is the fourth-order central 
moment. 
The Kurtosis $\mathcal{K}$, which 
measures the concentration 
of the distribution around its mean,
is an important
statistical quantity and 
widely
used in data analysis of astrophysics
and space physics. 
Using Equations 
(\ref{beta1204})
and (\ref{fourth order central moment
with alpha1101})
, we can obtain 
the Kurtosis formula 
\begin{eqnarray}
\mathcal{K}=\frac{\left\langle \left(z-\langle z \rangle \right)^4 \right\rangle}
{\left\langle \left(z-\langle z \rangle \right)^2 \right\rangle^2}
=\frac{\alpha^4_{1101}
+3\left(\alpha^1_{1101}\right)^2}
{\left(\alpha^2_{1101}\right)^2}. 
\end{eqnarray}

As demonstrated in 
Equations 
(\ref{ddt Delta z of equation1204})-(\ref{delta1204}),
there exists one-to-one
correspondence between 
the transport coefficients
$\kappa_{1z}$, $\kappa_{2z}$,
$\kappa_{4z}$
and the statistical quantitie
$\alpha^1_{1101}$, $\alpha^2_{1101}$,
$\alpha^4_{1101}$.

\subsection{The transport coefficient formulas and the corresponding statistical
quantities of the equation
with $T_2$ and $T_4$}

The hyperdiffusion
equation derived by 
\citet{MalkovSagdeev2015}
with $T_2$ and $T_4$,
is shown as
\begin{eqnarray}
\frac{\partial{F}}{\partial{t}}=
\kappa_{2z}\frac{\partial^2{F}}{\partial{z^2}}
+\kappa_{4z}\frac{\partial^4{F}}{\partial{z^4}}. 
\end{eqnarray}
It is straightforward to
derive the transport 
coefficient formulas, which is 
given by
\begin{eqnarray}
&&\kappa_{2z}
=\frac{1}{2}
\frac{\dee}{\dee t}
\alpha^2_{0101},
\label{ddt kzz for equation0204}
\\
&&\kappa_{4z}=\frac{1}{24}	\frac{\dee}{\dee t}
\alpha^4_{0101}
\label{ddt k4z for equation0204}
\end{eqnarray}
with
\begin{eqnarray}
&&\alpha^2_{0101}
=\left\langle z^2\right\rangle,\\
&&\alpha^4_{0101}
=\left\langle z^4\right\rangle
-3\left\langle z^2\right\rangle^2. 
\end{eqnarray}
Here, we note that 
$\alpha^2_{0101}$ is a relatively
new statistical quantity and 
the formula
$\alpha^2_{0101}=\alpha^2_{01}$
holds. 
Additionly, Equations
(\ref{ddt kzz for equation0204})-(\ref{ddt k4z for equation0204})
give the one-to-one correspondence
between transport coefficients
$\kappa_{2z}$, $\kappa_{4z}$
and statistical quantities
$\alpha^2_{0101}$, $\alpha^4_{0101}$, respectively.  
Moreover, the simplified type of
the kurtosis formula 
can be derived for 
symmetrical distribution
function
\begin{eqnarray}
\mathcal{K}=\frac{\alpha^4_{0101}}
{\left(\alpha^2_{0101}\right)^2}
=6\frac{\kappa_{4z}}{\kappa_{2z}^2}
\frac{1}{t}
=\frac{\left\langle z^4\right\rangle
}{\left\langle z^2\right\rangle^2}-3. 
\end{eqnarray}
For the kurtosis $\mathcal{K}$, 
the latter form is more familiar to the community.

\subsection{The transport coefficient formulas and the corresponding statistical
quantities of the equation
with $T_1$ and $T_4$}

The last STGE, which has
the terms $T_1$ and $T_4$,
is shown as  
\begin{eqnarray}
\frac{\partial{F}}{\partial{t}}=
\kappa_{1z}\frac{\partial{F}}
{\partial{z}}
+\kappa_{4z}\frac{\partial^4{F}}{\partial{z^4}}. 
\end{eqnarray}
The transport coefficients 
of the latter equation
can be derived as
\begin{eqnarray}
&&\kappa_{1z}=\frac{\dee}{\dee t}
\alpha^1_{1001},
\\
&&\kappa_{4z}=
\frac{1}{24}\frac{\dee}{\dee t}
\alpha^4_{1001}
\end{eqnarray}
with
\begin{eqnarray}
&&\alpha^1_{1001}=-\langle z\rangle,\\
&&\alpha^4_{1001}=
\left\langle z^4\right\rangle
-4\left\langle z^3\right\rangle\langle z\rangle
+3\left\langle z^2\right\rangle^2.
\end{eqnarray}
It can be seen easily that 
the transport coefficients
$\kappa_{1z}$ and $\kappa_{4z}$
are, respectively,  
the formulas for time 
derivative of
the statistical quantities
$\alpha^1_{1001}$ and 
$\alpha^4_{1001}$.

\section{The transport coefficients 
and statistical 
quantities of the fifth-order 
transport equation}
\label{The transport coefficients 
and statistical 
quantities of the fifth-order 
transport equation}

As demonstrated in Subsection
\ref{The simplified types
of the general spatial transport 
equation}, 
there are sixteen different
STGEs for the fifth-order spatial
transport equation.
All of the transport coefficient
formulas for the STGEs 
and the corresponding statistical
quantities are listed
in Table 
\ref{The transport 
coefficients  and 
statistical quantities of the 
fifth-order transport 
equation-table}. 
It is shown that
most of the transport coefficients
are determined by the 
corresponding statistical 
quantities. 
Meanwhile, we find that 
the higher order terms
in the STGEs 
do not have any influence on the 
relationship between 
the transport coefficients
of lower order derivative terms
and corresponding statistical
quantities.
In addition,
we find that the coefficient
$\kappa_{5z}$ in  
the STGE with $T_1$, $T_2$,
and $T_5$ does not
have the corresponding 
statistical quantity. 
In fact,
we also derive 
some transport coefficients
of the sixth-order spatial 
transport equation,
which are not listed in this paper, 
and 
find that some coefficients 
also do not
have any corresponding 
statistical quantities.
Therefore,  
some interesting problems
have arisen, 
as follows:
What kind of transport
coefficients 
do not have corresponding
statistical quantities?  
Why don't they have it? 
What are the conditions for these coefficients to meet?
Furthermore, 
a general formula satisfied 
by any transport 
coefficient for any order
transport equation should be provided.

\section{The subdiffusion terms and
nonlocal effect}

Here, we explore the physical
meaning of the subdiffusion terms
$T_n$ with $n=3,4,5,\cdots$. 
For the sake of simplicity, the
convection effect $T_1$ is
eliminated from the 
general spatial transport 
equation derived by \citet{wq2019}.
Thus, the general equation 
is rewritten as
\begin{eqnarray}
	\frac{\partial{F}}{\partial{t}}=
	\kappa_{2z}
	\frac{\partial^2{F}}{\partial{z^2}}
	+\kappa_{3z}
	\frac{\partial^3{F}}{\partial{z^3}}
	+\kappa_{4z}
	\frac{\partial^4{F}}{\partial{z^4}}
	+\kappa_{5z}
	\frac{\partial^5{F}}{\partial{z^5}}
	+\kappa_{6z}
	\frac{\partial^6{F}}{\partial{z^6}}
	+\cdots.
	\label{governing equation in real 
		space without convection term}
\end{eqnarray}
In fact, the latter
equation can be expressed in terms of the continous 
description 
\begin{eqnarray}  
	\frac{\partial{F}}{\partial{t}}=
	\frac{\partial{J}}{\partial{z}}
	\label{continous equation}
\end{eqnarray}
with
\begin{eqnarray}  
	J=\kappa_{2z}
	\frac{\partial{F}}{\partial{z}}
	+\kappa_{3z}
	\frac{\partial^2{F}}{\partial{z^2}}
	+\kappa_{4z}
	\frac{\partial^3{F}}{\partial{z^3}}
	+\kappa_{5z}
	\frac{\partial^4{F}}{\partial{z^4}}
	+\kappa_{6z}
	\frac{\partial^5{F}}{\partial{z^5}}
	+\cdots.
\end{eqnarray}
Performing a Fourier transform
on the latter formula gives
\begin{eqnarray}  
	\hat{J}=\hat{\lambda}(k)\cdot
	\hat{P}(k)
\end{eqnarray}
with
\begin{eqnarray}
	&&\hat{\lambda}(k)=\kappa_{2z}
	+\kappa_{3z}ik
	+\kappa_{4z}(ik)^2
	+\kappa_{5z}(ik)^3
	+\kappa_{6z}(ik)^4
	+\cdots,\\
	&&\hat{P}(k)=ik \hat{F}. 
\end{eqnarray}
Using an inverse Fourier transform,
we find
\begin{eqnarray}  
	J(z,t)=\frac{1}{2\pi}
	\int_{-\infty}^{+\infty}\dee
	k \hat{\lambda}(k)\cdot\hat{P}(k)
	e^{ikz}
	=\lambda(z)\ast F(z,t)=
	\int_{-\infty}^{+\infty}\dee z'
	\lambda(z'-z)F(z',t),
\end{eqnarray}
with 
\begin{eqnarray}
	\lambda(z)=\kappa_{2z}
	+\kappa_{3z}\delta(z)
	+\kappa_{4z}\delta(z)^{(2)}
	+\kappa_{5z}\delta(z)^{(3)}
	+\kappa_{6z}\delta(z)^{(4)}
	+\cdots. 
	\label{lambda with delta(z)}
\end{eqnarray}
Obviously, the variable $z$ only
occurs in the subdiffusion terms,
$\lambda(z)$. 
Inserting the latter equation into
Equation (\ref{continous equation})
yields
\begin{eqnarray}  
\frac{\partial{F}}{\partial{t}}=
\frac{\partial{}}{\partial{z}}
\int_{-\infty}^{+\infty}\dee z'
\lambda(z'-z)F(z',t). 
\end{eqnarray}
The integral on the right-hand side
of the latter equation denotes
spatial non-locality.
By using the method of 
Legendre polynomial expansion,
\citet{BianEA2017} explored
the nonlocal 
effect of particle transport
caused by
the Fokker-Planck equation. 
Here, we find that 
the subdiffusion terms $T_n$ of 
general 
spatial equation are related to
the spatial nonlocality.

\section{SUMMARY AND CONCLUSION}
\label{SUMMARY AND CONCLUSION}
The parallel transport coefficients,
such as the parallel diffusion
coefficient,  
in charged particle transport 
equations
are particularly important
in space plasma physics and
astrophysics. 
In this paper,  
all of the 
simplified equations belonging to
the first-, second-, third-,
fourth-, and fifth-order 
spatial transport equations
are provided. 
We find that
the $n$th-order spatial transport 
equation has $2^{n-1}$ 
different forms of the STGEs. 
For example, 
the first-order ($n=1$) spatial 
transport equation
has only one form of the STGE,
i.e., the convection equation,
and the second-order ($n=2$) 
spatial transport 
equation has two forms of the
STGEs,
namely,
the convection  
and convection-diffusion
equations. 
The third-, fourth-, and fifth-order
spatial transport equations
have four, eight and 
sixteen different forms of the
STGEs, 
respectively. 
The hyperdiffusion and
subdiffusion transport equations
derived by \citet{MalkovSagdeev2015}
and 
\citet{ShalchiArendt2020},
respectively, 
belong to
the fourth-order 
spatial transport equation. 

In this article, 
all of the transport 
coefficient formulas
for the first-, second-, third-, fourth-, and fifth-order 
STGEs
are derived. 
It is shown
that most of the transport 
coefficients 
are determined by the corresponding 
statistical quantities. 
For example, 
the convection coefficient
$\kappa_{1z}$ is determined by
the mathematical expectation 
$\langle z\rangle$,
the diffusion coefficient $\kappa_{2z}$
in the convection-diffusion 
equation by the variance, i.e.,
$\left\langle z^2 \right\rangle
-\langle  z\rangle^2
=\left\langle \left(z
-\langle z \rangle \right)^2 \right\rangle$, 
the third-order transport
coefficient $\kappa_{3z}$
in the STGE with $T_1$, $T_2$
and $T_3$
by the third-order central 
moment of the charged particle
distribution function,
namely, $\left\langle \left(z
-\langle z \rangle \right)^3 \right\rangle$, 
the fourth-order transport
coefficient $\kappa_{4z}$
in the STGE with $T_1$, $T_3$
and $T_4$
by the fourth-order central 
moment of distribution function,
namely, $\left\langle \left(z
-\langle z \rangle \right)^4 \right\rangle$, 
etc. 
Additionly, it is shown that
the higher spatial derivative
terms do not influence 
the relationship between
transport coefficients
and corresponding statistical
quantities. 
Meanwhile, we identify 
a number of statistical quantities
that 
are relatively new and could 
be important in
some scenarios. 
These statistical quantities
need to be further investigated 
in future studies. 
In addition,  
we find that subdiffusion
terms should be related to 
spatial nonlocality. 
We will explore this problem. 

Moreover, we find that 
the skewness $\mathcal{S}$, 
which describes the 
uniformity of the distribution
function, can be expressed 
by the statistical quantities
$\alpha^2_{111}$ and $\alpha^3_{111}$,
i.e., $\mathcal{S}$ is determined
by the transport coefficients
$\kappa_{2z}$ and $\kappa_{3z}$ 
in the STGE with $T_1$, $T_2$
and $T_3$.
In addition, the kurtosis
$\mathcal{K}$, which
measures the concentration 
of the distribution around its mean,
can be expressed by the statistical
quantities $\alpha^1_{1101}$,
$\alpha^2_{1101}$, and $\alpha^4_{1101}$.
In other words, the kurtosis
$\mathcal{K}$ is determined 
by the transport
coefficients $\kappa_{1z}$,
$\kappa_{2z}$, and $\kappa_{4z}$
in the STGE with $T_1$, $T_2$
and $T_4$.
It is demonstrated that 
these important 
statistical quantities 
are related
to subdiffusion processes and 
are determined by the subdiffusive
coeffcients of certain 
transport equations. 
This is an interesting discovery. 
In the future, 
with these partial
differential subdiffusion
equations, 
some further understanding of these important
statistical quantities 
might be achieved. 

In addition, the parallel 
transport coefficients 
are the important parameters 
for particle
shock acceleration, 
the solar modulation of 
cosmic rays and so on. 
\citet{Shalchi2016}
explored
the implicit contribution
of the subdiffusion to perpendicular
transport coefficient and found 
that,
for some cases, 
this implicit perpendicular 
subdiffusion 
contribution can have a 
significant effect
on the transport coefficient.
Accordingly, the effect of 
perpendicular subdiffusion 
should have 
an important
influence on particle shock 
acceleration, the modulation of
cosmic rays, and so on. 
Similarly, as the important 
input parameter, the parallel
transport coefficients with 
parallel subdiffusion effect 
should also be throughly 
explored.
This is also part of our future work.

Moreover, we find that
a few transport
coefficient formulas do not have 
corresponding statistical 
quantities, e.g.,   
$\kappa_{5z}$ in 
the STGE
with $T_1$, $T_2$, and $T_5$. 
What's more, there are some 
problems that need to be 
explored: 
why do some coefficients
not have corresponding
statistical quantities? 
What types of coefficients 
do not have the corresponding
statistical quantities?  
What are the conditions 
that these coefficients need 
to satisfy?
In addition, a general formula
that is satisfied by any transport coefficient for any order 
transport equation should 
be provided. 
In the future,
we will explore these problems 
further.
This work can help 
one to use different transport
coefficients, which are
determined by
the statistical quantities, 
including 
many that are relatively new 
found in this paper, 
to study charged particle
parallel transport processes.


\begin{acknowledgments}
The authors thank the anonymous referee for their valuable
comments.
We are partly supported by the Shenzhen Science and Technology 
Program under Grant No. JCYJ20210324132812029, and by
grants NNSFC 42074206 and NNSFC 42150105.
The work was supported by the National Key R\&D program of China 
No.2021YFA0718600, and by Shenzhen Key
Laboratory Launching Project (No. ZDSYS20210702140800001). 
\end{acknowledgments}

\renewcommand{\theequation}{\Alph{section}-\arabic{equation}}
\setcounter{equation}{0}  

\begin{table}[ht]
	
\begin{center}
	\caption{The transport coefficients and 
		statistical quantities of the 
		third-order transport equation}
	\label{The transport coefficients and 
		statistical quantities of the 
	third-order transport equation-Table}    	
	\begin{tabular}{|l|l|l|l|} 
		\hline
		\multirow{1}{*}{Transport equations} & Transport
		coefficients & Statistical
		quantities \\
		\hline    
		\multirow{3}{*}{$\frac{\partial{F}}{\partial{t}}
			=
			\kappa_{1z}\frac{\partial{F}}
			{\partial{z}}
			+\kappa_{2z}\frac{\partial^2{F}}
			{\partial{z^2}}
			+\kappa_{3z}\frac{\partial^3{F}}
			{\partial{z^3}}$} 

& $\kappa_{1z}=
\frac{\dee}{\dee t}
\alpha^1_{111}$ & $\alpha^1_{111}=-\langle z\rangle$ \\
\cline{2-3} 
& $\kappa_{2z}=
\frac{1}{2}\frac{\dee}{\dee t}
\alpha^2_{111}$ & $\alpha^2_{111}=\left\langle z^2 \right\rangle
-\langle z\rangle^2$ \\
\cline{2-3} 			
& $\kappa_{3z}=\frac{1}{6}
\frac{\dee}{\dee t}\alpha^3_{111}$ & $\alpha^3_{111}=
3\left\langle z \right\rangle\left\langle z^2 \right\rangle
-2\left\langle z \right\rangle^3
-\left\langle z^3 \right\rangle$ \\		
\cline{2-3} 								
\hline

		\multirow{1}{*}{$\frac{\partial{F}}{\partial{t}}
			=
			\kappa_{3z}\frac{\partial^3{F}}
			{\partial{z^3}}$} 

& $\kappa_{3z}=\frac{1}{6}
\frac{\dee}{\dee t}
\alpha^3_{001}$ & $\alpha^3_{001}
=-\left\langle z^3 \right\rangle$ \\
\cline{2-3} 								
\hline

\multirow{2}{*}{$\frac{\partial{F}}{\partial{t}}=
	\kappa_{1z}\frac{\partial{F}}
	{\partial{z}}
	+\kappa_{3z}\frac{\partial^3{F}}{\partial{z^3}}$} 

& $\kappa_{1z}=\frac{\dee}{\dee t}
\alpha^1_{101}$ & $\alpha^1_{101}=
-\left\langle z\right\rangle$ \\
\cline{2-3} 								
& $\kappa_{3z}=\frac{1}{6}
\frac{\dee}{\dee t}
\alpha^3_{101}$ & $\alpha^3_{101}=3\langle z\rangle
\left\langle z^2 \right\rangle
-2\left\langle z 
\right\rangle^3
-\left\langle z^3 \right\rangle$ \\
\cline{2-3} 								
\hline

\multirow{2}{*}{$\frac{\partial{F}}{\partial{t}}
	=
	\kappa_{2z}\frac{\partial^2{F}}
	{\partial{z^2}}
	+\kappa_{3z}\frac{\partial^3{F}}
	{\partial{z^3}}$} 

& $\kappa_{2z}=
\frac{1}{2}\frac{\dee}{\dee t}
\alpha^2_{011}$ & 
$\alpha^2_{011}=\left\langle z^2 \right\rangle$ \\
\cline{2-3} 								
& $\kappa_{3z}=\frac{1}{6}	\frac{d}{dt}\alpha^3_{011}$ & 
$\alpha^3_{011}=
3\left\langle z \right\rangle
\left\langle z^2 \right\rangle
-\left\langle z^3 \right\rangle$ \\
\cline{2-3} 								
\hline

\end{tabular}        	   
\end{center}    

\end{table}

\begin{table}[ht]
	
\begin{center}
	\caption{The transport coefficients and 
		statistical quantities of the 
		fourth-order transport equation}
	\label{The transport coefficients and statistical quantities of the 
	fourth-order transport equation-Table}    	
	\begin{tabular}{|l|l|l|l|} 
		\hline  
		\multirow{1}{*}{Transport equations} & Transport
		coefficients & Statistical
		quantities \\
		\hline       	
		\multirow{4}{*}{$\frac{\partial{F}}{\partial{t}}=
			\kappa_{1z}\frac{\partial{F}}
			{\partial{z}}
			+\kappa_{2z}\frac{\partial^2{F}}{\partial{z^2}}
			+\kappa_{3z}\frac{\partial^3{F}}{\partial{z^3}}
			+\kappa_{4z}\frac{\partial^4{F}}{\partial{z^4}}$} 
		
		& $\kappa_{1z}=\frac{\dee}{\dee t}
		\alpha^1_{1111}$ & $\alpha^1_{1111}=-\langle z\rangle$ \\
		\cline{2-3} 
		& $\kappa_{2z}=
		\frac{1}{2}\frac{\dee}{\dee t}
		\alpha^2_{1111}$ & $\alpha^2_{1111}=\left\langle z^2\right\rangle
		-\langle z\rangle^2$ \\
		\cline{2-3} 			
		& $\kappa_{3z}=	\frac{1}{6}\frac{\dee}{\dee t}
		\alpha^3_{1111}$ & $\alpha^3_{1111}=
		3\left\langle z \right\rangle\left\langle z^2 \right\rangle
		-\left\langle z^3\right\rangle
		-2\left\langle z \right\rangle^3$ \\		
		\cline{2-3} 			
		& $\kappa_{4z}=\frac{1}{24}\frac{d}{dt}
		\alpha^4_{1111}$ & $\alpha^4_{1111}=\left\langle z^4
		\right\rangle
		+12
		\left\langle z^2\right\rangle\langle z\rangle^2
		-4
		\langle z\rangle\left\langle z^3\right\rangle
		-3\left\langle z^2\right\rangle^2
		-6\left\langle z \right\rangle^4$ \\			
		\cline{2-3} 					
		\hline

		\multirow{4}{*}{$\frac{\partial{F}}{\partial{t}}=
			\kappa_{2z}\frac{\partial^2{F}}{\partial{z^2}}
			+\kappa_{3z}\frac{\partial^3{F}}{\partial{z^3}}
			+\kappa_{4z}\frac{\partial^4{F}}{\partial{z^4}}
			\label{234}$} 

& $\kappa_{2z}=\frac{1}{2}
\frac{\dee}{\dee t}
\alpha^2_{0111}$ & $\alpha^2_{0111}=	
\left\langle z^2\right\rangle$ \\
\cline{2-3} 
& $\kappa_{3z}=
\frac{1}{6}\frac{\dee}{\dee t}
\alpha^3_{0111}$ & $\alpha^3_{0111}=3\left\langle z \right\rangle
\left\langle z^2 \right\rangle 
-\left\langle z^3 \right\rangle$ \\
\cline{2-3} 			
& $\kappa_{4z}=\frac{1}{24}
\frac{d}{dt}\alpha^4_{0111}$ & $\alpha^4_{0111}=
\left\langle z^4\right\rangle
-3\left\langle z^2\right\rangle^2
-4\langle z\rangle 
\left\langle z^3\right\rangle
+12\langle z\rangle^2
\left\langle z^2\right\rangle$ \\		
\cline{2-3} 			 					
\hline

\multirow{2}{*}{$\frac{\partial{F}}{\partial{t}}=
	\kappa_{3z}\frac{\partial^3{F}}{\partial{z^3}}
	+\kappa_{4z}\frac{\partial^4{F}}{\partial{z^4}}$} 

& $\kappa_{3z}=
\frac{1}{6}\frac{\dee}{\dee t}
\alpha^3_{0011}$ & $\alpha^3_{0011}=
-\left\langle z^3\right\rangle$ \\
\cline{2-3} 
& $\kappa_{4z}=\frac{1}{24}
\frac{\dee}{\dee t}
\alpha^4_{0011}$ & $\alpha^4_{0011}
=\left\langle z^4\right\rangle
-4\langle z\rangle 
\left\langle z^3\right\rangle$ \\
\cline{2-3}			 					
\hline

\multirow{1}{*}{$\frac{\partial{F}}{\partial{t}}=
	\kappa_{4z}\frac{\partial^4{F}}{\partial{z^4}}$} 

& $\kappa_{4z}=\frac{1}{24}
\frac{\dee}{\dee t}
\alpha^4_{0001}$ & $\alpha^4_{0001}
=\left\langle z^4\right\rangle$ \\
\cline{2-3} 		 					
\hline

\multirow{3}{*}{$\frac{\partial{F}}{\partial{t}}=
	\kappa_{1z}\frac{\partial{F}}
	{\partial{z}}
	+\kappa_{3z}\frac{\partial^3{F}}{\partial{z^3}}
	+\kappa_{4z}\frac{\partial^4{F}}{\partial{z^4}}$} 

& $\kappa_{1z}=
\frac{\dee}{\dee t}\alpha^1_{1011}$ & $\alpha^1_{1011}=-\langle z\rangle$ \\
\cline{2-3} 
& $\kappa_{3z}=
\frac{1}{6}\frac{\dee}{\dee t}
\alpha^3_{1011}$ & $\alpha^3_{1011}=3\langle z\rangle
\left\langle z^2 \right\rangle
-2\left\langle z 
\right\rangle^3
-\left\langle z^3 \right\rangle$ \\
\cline{2-3}	
& $\kappa_{4z}
=\frac{1}{24}\frac{\dee}{\dee t}
\alpha^4_{1011}$ & $\alpha^4_{1011}=
\left\langle z^4 \right\rangle
-4\left\langle z^3\right\rangle\langle z\rangle
+6\left\langle z^2 \right\rangle\langle z\rangle^2
-3
\left\langle z \right\rangle^4$ \\
\cline{2-3}			 					
\hline	

\multirow{3}{*}{$\frac{\partial{F}}{\partial{t}}=
	\kappa_{1z}\frac{\partial{F}}
	{\partial{z}}
	+\kappa_{2z}\frac{\partial^2{F}}{\partial{z^2}}
	+\kappa_{4z}\frac{\partial^4{F}}{\partial{z^4}}$} 

& $\kappa_{1z}=
\frac{d}{dt}\alpha^1_{1101}$ & $\alpha^1_{1101}=-\langle z\rangle$ \\
\cline{2-3} 
& $\kappa_{2z}=
\frac{1}{2}\frac{\dee}{\dee t}
\alpha^2_{1101}$ & $\alpha^2_{1101}=\left\langle z^2\right\rangle
-\langle z\rangle^2$ \\
\cline{2-3}	
& $\kappa_{4z}
=\frac{1}{24}\frac{\dee}{\dee t}
\alpha^4_{1101}$ & $\alpha^4_{1101}=
\langle z^4\rangle
-4\langle z\rangle\left\langle z^3\right\rangle
+12\left\langle z^2 \right\rangle\langle z\rangle^2
-3\left\langle z^2\right\rangle^2
-6\langle z\rangle^4$ \\
\cline{2-3}			 					
\hline	

\multirow{2}{*}{$\frac{\partial{F}}{\partial{t}}=
	\kappa_{2z}\frac{\partial^2{F}}{\partial{z^2}}
	+\kappa_{4z}\frac{\partial^4{F}}{\partial{z^4}}$} 

& $\kappa_{2z}
=\frac{1}{2}
\frac{\dee}{\dee t}
\alpha^2_{0101}$ & $\alpha^2_{0101}
=\left\langle z^2\right\rangle$ \\
\cline{2-3} 
& $\kappa_{4z}=\frac{1}{24}	\frac{\dee}{\dee t}
\alpha^4_{0101}$ & $\alpha^4_{0101}
=\left\langle z^4\right\rangle
-3\left\langle z^2\right\rangle^2$ \\
\cline{2-3}			 					
\hline	

\multirow{2}{*}{$\frac{\partial{F}}{\partial{t}}=
	\kappa_{1z}\frac{\partial{F}}
	{\partial{z}}
	+\kappa_{4z}\frac{\partial^4{F}}{\partial{z^4}}$} 

& $\kappa_{1z}=\frac{\dee}{\dee t}
\alpha^1_{1001}$ & $\alpha^1_{1001}=
-\langle z\rangle$ \\
\cline{2-3} 
& $\kappa_{4z}=
\frac{1}{24}\frac{\dee}{\dee t}
\alpha^4_{1001}$ & $\alpha^4_{1001}=
\left\langle z^4\right\rangle
-4\left\langle z^3\right\rangle\langle z\rangle
+3\left\langle z^2\right\rangle^2$ \\
\cline{2-3}			 					
\hline

\end{tabular}        	   
\end{center}    

\end{table}

\begin{table}[ht]

    \begin{center}
		\caption{The transport 
			coefficients  and 
			statistical quantities of the 
			fifth-order transport equation}
        \label{The transport 
        	coefficients  and 
        	statistical quantities of the 
        	fifth-order transport equation-table}    	
        \begin{tabular}{|l|l|l|l|} 
			\hline
			\multirow{1}{*}{Transport equations} & Transport
			coefficients & Statistical
			quantities \\
			\hline         	
			\multirow{7}{*}{$\frac{\partial{F}}{\partial{t}}
			=\kappa_{1z}\frac{\partial{F}}{\partial{z}}
			+\kappa_{2z}\frac{\partial^2{F}}{\partial{z^2}}+\kappa_{3z}\frac{\partial^3{F}}{\partial{z^3}}+\kappa_{4z}\frac{\partial^4{F}}{\partial{z^4}}+\kappa_{5z}\frac{\partial^5{F}}{\partial{z^5}}$
			} 
			
			& $\kappa_{1z}=\frac{\dee}{\dee t}\alpha^1_{11111}$ & $\alpha^1_{11111}=-\langle z\rangle$ \\
			\cline{2-3} 
			& $\kappa_{2z}=\frac{1}{2}
			\frac{\dee}{\dee t}
			\alpha^2_{11111}$ & $\alpha^2_{11111}=
			\left\langle z^2\right\rangle
			-\langle z\rangle^2$ \\
			\cline{2-3} 			
            & $\kappa_{3z}
            =\frac{1}{6}
            \frac{\dee}{\dee t}
            \alpha^3_{11111}$ & $\alpha^3_{11111}=3\left\langle z\right\rangle\left\langle z^2 \right\rangle
            -\left\langle z^3\right\rangle
            -2\langle z\rangle^3$ \\		
			\cline{2-3} 			
            & $\kappa_{4z}=\frac{1}{24}
            \frac{\dee}{\dee t}
            \alpha^4_{11111}$ & $\alpha^4_{11111}=
            \left\langle z^4
            \right\rangle
            +12
            \left\langle z^2\right\rangle\langle z\rangle^2
            -4
            \langle z\rangle\left\langle z^3\right\rangle
            -3\left\langle z^2\right\rangle^2$\\
            & &
            $-6\left\langle z \right\rangle^4$ \\			
			\cline{2-3} 			
            & $\kappa_{5z}=\frac{1}{120}
             \frac{\dee}{\dee t}
             \alpha^5_{11111}$ 
            & $\alpha^5_{11111}
             =-\left\langle z^5\right\rangle
             +10\left\langle  z^3\right\rangle
             \left\langle z^2\right\rangle
             +5\left\langle  
             z^4\right\rangle\langle z\rangle$\\
             & &
             $
             -24\langle z\rangle^5-20\left\langle  
             z^3\right\rangle
             \langle z\rangle^2
            -30\left\langle  
            z\right\rangle\left\langle 
            z^2\right\rangle^2
            $\\
            & &
            $
            +60\left\langle  z^2\right\rangle\langle 
            z\rangle^3$ 
            \\				
			\hline
			\multirow{3}{*}{$\frac{\partial{F}}{\partial{t}}=
	        \kappa_{1z}\frac{\partial{F}}
	         {\partial{z}}
	         +\kappa_{5z}
	         \frac{\partial^5{F}}
	         {\partial{z^5}}$} 

             & $\kappa_{1z}=\frac{\dee}{\dee t}
              \alpha^1_{10001}$ 
             & $\alpha^1_{10001}=-\langle 
              z\rangle$ \\
             \cline{2-3} 			

& $\kappa_{5z}
=\frac{1}{120}\frac{\dee}{\dee t}
\alpha^5_{10001}$ 
& $\alpha^5_{10001}=
5\left\langle  z^4
\right\rangle\langle z\rangle
-10\left\langle z^3
\right\rangle\langle z
\rangle^2
+10\left\langle z^2 \right\rangle\langle z\rangle^3$\\
& &
$-4
\left\langle z \right\rangle^5 
-\left\langle z^5 \right\rangle$ \\
\cline{2-3} 				
			\hline				
			
\multirow{3}{*}{$\frac{\partial{F}}{\partial{t}}=
	\kappa_{2z}\frac{\partial^2{F}}
	{\partial{z^2}}
	+\kappa_{5z}\frac{\partial^5{F}}
	{\partial{z^5}}$} 

& $\kappa_{2z}=
\frac{1}{2}\frac{\dee}{\dee t}
\alpha^2_{01001}$ 
& $\alpha^2_{01001}=\left\langle z^2\right\rangle$ \\
\cline{2-3} 			

& $\kappa_{5z}
=\frac{1}{120}\frac{\dee}{\dee t}
\alpha^5_{01001}$ 
& $\alpha^5_{01001}=
10\left\langle  z^3\right\rangle
\left\langle  z^2\right\rangle
-15\left\langle z \right\rangle
\left\langle  z^2
\right\rangle^2
-\left\langle z^5\right\rangle$ \\
\cline{2-3} 				
\hline				
			
\multirow{3}{*}{$\frac{\partial{F}}{\partial{t}}=
	\kappa_{3z}\frac{\partial^3{F}}{\partial{z^3}}
	+\kappa_{5z}\frac{\partial^5{F}}{\partial{z^5}}$} 

& $\kappa_{3z}=\frac{1}{6}
\frac{\dee}{\dee t}\alpha^3_{00101}$ 
& $\alpha^3_{00101}=
-\left\langle z^3 \right\rangle$ \\
\cline{2-3} 			

& $\kappa_{5z}
=\frac{1}{120}\frac{\dee}{\dee t}
\alpha^5_{00101}$ 
& $\alpha^5_{00101}=10\left\langle  z^2\right\rangle
\left\langle z^3 \right\rangle
-\left\langle z^5 \right\rangle$ \\
\cline{2-3} 				
\hline							
						
\multirow{3}{*}{$\frac{\partial{F}}{\partial{t}}=
	\kappa_{4z}\frac{\partial^4{F}}{\partial{z^4}}
	+\kappa_{5z}\frac{\partial^5{F}}{\partial{z^5}}$} 

& $\kappa_{4z}=\frac{1}{24}
\frac{\dee}{\dee t}
\alpha^4_{00011}$ 
& $\alpha^4_{00011}=
\left\langle z^4 \right\rangle$ \\
\cline{2-3} 			

& $\kappa_{5z}=\frac{1}{120}
\frac{\dee}{\dee t}
\alpha^5_{00011}$ 
& $\alpha^5_{00011}=5\left\langle  z\right\rangle
\left\langle z^4 \right\rangle
-\left\langle z^5 \right\rangle$ \\
\cline{2-3} 				
\hline				

\multirow{3}{*}{$\frac{\partial{F}}{\partial{t}}=
	\kappa_{1z}\frac{\partial{F}}
	{\partial{z}}
	+\kappa_{2z}\frac{\partial^2{F}}{\partial{z^2}}
	+\kappa_{5z}\frac{\partial^5{F}}{\partial{z^5}}$} 

& $\kappa_{1z}=\frac{\dee}{\dee t}
\alpha^1_{11001}$ 
& $\alpha^1_{11001}=-\langle z\rangle$ \\
\cline{2-3} 			

& $\kappa_{2z}=
\frac{1}{2}\frac{\dee}{\dee t}
\alpha^2_{11001}$ 
& $\alpha^2_{11001}=\left\langle z^2
\right\rangle
-\langle z\rangle^2$ \\
\cline{2-3}

& $\kappa_{5z}$ is not existent 
& No exist \\
\cline{2-3} 				
\hline		

\multirow{4}{*}{$\frac{\partial{F}}{\partial{t}}=
	\kappa_{1z}\frac{\partial{F}}
	{\partial{z}}
	+\kappa_{3z}\frac{\partial^3{F}}{\partial{z^3}}
	+\kappa_{5z}\frac{\partial^5{F}}{\partial{z^5}}$} 

& $\kappa_{1z}=\frac{\dee}{\dee t}\alpha^1_{10101}$ 
& $\alpha^1_{10101}=-\langle z\rangle$ \\
\cline{2-3} 			

& $\kappa_{3z}=\frac{1}{6}
\frac{\dee}{\dee t}
\alpha^3_{10101}$ 
& $\alpha^3_{10101}=3\langle z\rangle
\left\langle z^2 \right\rangle
-2\left\langle z 
\right\rangle^3
-\left\langle z^3 \right\rangle$ \\
\cline{2-3}

& $\kappa_{5z}=\frac{1}{120}
\frac{\dee}{\dee t}
\alpha^5_{10101}$ 
& $\alpha^5_{10101}=
5\left\langle  z^4
\right\rangle
\langle z\rangle
-44
\left\langle z 
\right\rangle^5
-20
\left\langle z^3\right\rangle
\left\langle z\right\rangle^2
$\\
& &
$
+10\left\langle  
z^2\right\rangle
\left\langle z^3 \right\rangle+80\langle z\rangle^3
\left\langle z^2 \right\rangle
-30\left\langle  
z^2\right\rangle^2
\langle z\rangle
-\left\langle z^5\right\rangle$ \\
\cline{2-3} 				
\hline		

\multirow{4}{*}{$\frac{\partial{F}}{\partial{t}}=
	\kappa_{1z}\frac{\partial{F}}
	{\partial{z}}
	+\kappa_{4z}\frac{\partial^4{F}}{\partial{z^4}}
	+\kappa_{5z}\frac{\partial^5{F}}{\partial{z^5}}$} 

& $\kappa_{1z}=\frac{\dee}{\dee t}
\alpha^1_{10011}$ 
& $\alpha^1_{10011}=-\langle z\rangle$ \\
\cline{2-3} 			

& $\kappa_{4z}=
\frac{1}{24}\frac{\dee}{\dee t}
\alpha^4_{10011}$ 
& $\alpha^4_{10011}=
\left\langle z^4 \right\rangle
-4\left\langle z^3\right\rangle
\langle z\rangle
+3\left\langle z^2 \right\rangle^2$ \\
\cline{2-3}

& $\kappa_{5z}=\frac{1}{120}
\frac{\dee}{\dee t}\alpha^5_{10011}$ 
& $\alpha^5_{10011}=
-\left\langle z^5 \right\rangle
+5 \left\langle  z^4\right\rangle
\langle z\rangle
-10\left\langle z^3\right\rangle
\langle z\rangle^2$\\
& &
$+10
\left\langle z^2 \right\rangle
\langle z\rangle^3
-4\left\langle z \right\rangle^5$ \\
\cline{2-3} 				
\hline

		\end{tabular}        	   
	\end{center}    

\end{table}

\begin{table}[ht]
	\begin{center}

		\begin{tabular}{|l|l|l|l|}
			\hline

\multirow{4}{*}{$\frac{\partial{F}}{\partial{t}}=
	\kappa_{2z}\frac{\partial^2{F}}{\partial{z^2}}
	+\kappa_{3z}\frac{\partial^3{F}}{\partial{z^3}}
	+\kappa_{5z}\frac{\partial^5{F}}{\partial{z^5}}$} 

& $\kappa_{2z}=
\frac{1}{2}\frac{\dee}{\dee t}
\alpha^2_{01101}$ 
& $\alpha^2_{01101}=
\left\langle z^2 \right\rangle$ \\
\cline{2-3} 			

& $\kappa_{3z}=\frac{1}{6}	
\frac{\dee}{\dee t}\alpha^3_{01101}$ 
& $\alpha^3_{01101}=
3\left\langle z \right\rangle
\left\langle z^2 \right\rangle
-\left\langle z^3 \right\rangle$ \\
\cline{2-3}

& $\kappa_{5z}=\frac{1}{120}
\frac{\dee}{\dee t}\alpha^5_{01101}$ 
& $\alpha^5_{01101}=
-\left\langle z^5 \right\rangle
+10\left\langle  z^3\right\rangle
\left\langle  z^2\right\rangle
-30\left\langle z \right\rangle
\left\langle z^2\right\rangle^2
+5\left\langle  z^4\right\rangle
\left\langle  z\right\rangle$ \\
& &
$-20
\left\langle z^3\right\rangle
\left\langle  z\right\rangle^2
+60
\left\langle  z\right\rangle^3
\left\langle z^2 \right\rangle$ \\
\cline{2-3} 				
\hline

\multirow{4}{*}{$\frac{\partial{F}}{\partial{t}}=
	\kappa_{2z}\frac{\partial^2{F}}{\partial{z^2}}
	+\kappa_{4z}\frac{\partial^4{F}}{\partial{z^4}}
	+\kappa_{5z}\frac{\partial^5{F}}{\partial{z^5}}$} 

& $\kappa_{2z}=\frac{1}{2}
\frac{\dee}{\dee t}\alpha^2_{01011}$ 
& $\alpha^2_{01011}=
\left\langle z^2\right\rangle$ \\
\cline{2-3} 			

& $\kappa_{4z}=
\frac{1}{24}\frac{\dee}{\dee t}
\alpha^4_{01011}$ 
& $\alpha^4_{01011}=
\left\langle z^4\right\rangle
-3\left\langle z^2\right\rangle^2$ \\
\cline{2-3}

& $\kappa_{5z}=\frac{1}{120}
\frac{\dee}{\dee t}\alpha^5_{01011}$ 
& $\alpha^5_{01011}=
-\left\langle z^5 \right\rangle
+10\left\langle  z^3\right\rangle \left\langle z^2 \right\rangle
-15\left\langle z \right\rangle
\left\langle z^2 \right\rangle^2
+5\left\langle  z\right\rangle
\left\langle z^4 \right\rangle$\\
& &
$-15\left\langle  z\right\rangle
\left\langle z^2 \right\rangle^2$ \\
\cline{2-3} 				
\hline		

\multirow{4}{*}{$\frac{\partial{F}}{\partial{t}}=
	\kappa_{3z}\frac{\partial^3{F}}{\partial{z^3}}
	+\kappa_{4z}\frac{\partial^4{F}}{\partial{z^4}}
	+\kappa_{5z}\frac{\partial^5{F}}{\partial{z^5}}$} 

& $\kappa_{3z}=
\frac{1}{6}\frac{\dee}{\dee t}
\alpha^3_{00111}$ 
& $\alpha^3_{00111}=
-\left\langle z^3\right\rangle$ \\
\cline{2-3} 			

& $\kappa_{4z}=\frac{1}{24}
\frac{\dee}{\dee t}
\alpha^4_{00111}$ 
& $\alpha^4_{00111}
=\left\langle z^4\right\rangle
-4\langle z\rangle 
\left\langle z^3\right\rangle$ \\
\cline{2-3}

& $\kappa_{5z}=\frac{1}{120}
\frac{\dee}{\dee t}
\alpha^5_{00111}$ 
& $\alpha^5_{00111} 
=-\left\langle z^5 \right\rangle
+5\left\langle  z\right\rangle 
\left\langle z^4 \right\rangle
-20\left\langle  z\right\rangle^2
\left\langle z^3 \right\rangle
+10\left\langle  z^2\right\rangle
\left\langle z^3 \right\rangle$ \\
\cline{2-3} 				
\hline		

\multirow{5}{*}{$\frac{\partial{F}}{\partial{t}}=
	\kappa_{1z}\frac{\partial{F}}
	{\partial{z}}
	+\kappa_{3z}\frac{\partial^3{F}}{\partial{z^3}}
	+\kappa_{4z}\frac{\partial^4{F}}{\partial{z^4}}
	+\kappa_{5z}\frac{\partial^5{F}}{\partial{z^5}}$} 

& $\kappa_{1z}=
\frac{\dee}{\dee t}\alpha^1_{10111}$ 
& $\alpha^1_{10111}=-\langle z\rangle$ \\
\cline{2-3} 			

& $\kappa_{3z}=
\frac{1}{6}\frac{\dee}{\dee t}
\alpha^3_{10111}$ 
& $\alpha^3_{10111}=3\langle z\rangle
\left\langle z^2 \right\rangle
-2\left\langle z 
\right\rangle^3
-\left\langle z^3 \right\rangle$ \\
\cline{2-3}

& $\kappa_{4z}
=\frac{1}{24}\frac{\dee}{\dee t}
\alpha^4_{10111}$ 
& $\alpha^4_{10111}=
\left\langle z^4 \right\rangle
-4\left\langle z^3\right\rangle\langle z\rangle
+6\left\langle z^2 \right\rangle\langle z\rangle^2
-3
\left\langle z \right\rangle^4$ \\
\cline{2-3} 					

& $\kappa_{5z}
=\frac{1}{120}\frac{\dee}{\dee t}
\alpha^5_{10111}$ 
& $\alpha^5_{10111}=
-\left\langle z^5 \right\rangle
+5\left\langle  z^4
\right\rangle\langle z\rangle
-20\left\langle z^3
\right\rangle\langle z\rangle^2
+60\left\langle z^2 \right\rangle
\langle z\rangle^3$\\
& &
$+10\left\langle z^3 \right\rangle
\left\langle z^2 \right\rangle
-30\left\langle  z^2
\right\rangle^2
\langle z\rangle
-24\left\langle  z\right\rangle^5$ \\
\cline{2-3} 				
\hline

\multirow{5}{*}{$\frac{\partial{F}}{\partial{t}}=
	\kappa_{1z}\frac{\partial{F}}
	{\partial{z}}
	+\kappa_{2z}\frac{\partial^2{F}}{\partial{z^2}}
	+\kappa_{4z}\frac{\partial^4{F}}{\partial{z^4}}
	+\kappa_{5z}\frac{\partial^5{F}}{\partial{z^5}}$} 

& $\kappa_{1z}=
\frac{\dee}{\dee t}\alpha^1_{11011}$ 
& $\alpha^1_{11011}=-\langle z\rangle$ \\
\cline{2-3} 			

& $\kappa_{2z}=
\frac{1}{2}\frac{\dee}{\dee t}
\alpha^2_{11011}$ 
& $\alpha^2_{11011}=\left\langle z^2\right\rangle
-\langle z\rangle^2$ \\
\cline{2-3}

& $\kappa_{4z}
=\frac{1}{24}\frac{\dee}{\dee t}
\alpha^4_{11011}$ 
& $\alpha^4_{11011}=
\left\langle z^4 \right\rangle
-4\langle z\rangle\left\langle z^3\right\rangle
+12\left\langle z^2 \right\rangle\langle z\rangle^2
-3\left\langle z^2 \right\rangle^2
-6\langle z\rangle^4$ \\
\cline{2-3} 					

& $\kappa_{5z}
=\frac{1}{120}\frac{\dee}{\dee t}
\alpha^5_{11011}$ 
& $\alpha^5_{11011}=
-\left\langle z^5 \right\rangle
+5\left\langle  z^4
\right\rangle\langle z\rangle 
-10\left\langle z^3
\right\rangle\langle z\rangle^2
+20\left\langle z^2 
\right\rangle\langle z\rangle^3$\\
& &
$-8\langle z\rangle^5$ \\
\cline{2-3} 	
\hline

\multirow{5}{*}{$\frac{\partial{F}}{\partial{t}}=
	\kappa_{2z}\frac{\partial^2{F}}{\partial{z^2}}
	+\kappa_{3z}\frac{\partial^3{F}}{\partial{z^3}}
	+\kappa_{4z}\frac{\partial^4{F}}{\partial{z^4}}
	+\kappa_{5z}\frac{\partial^5{F}}{\partial{z^5}}$} 

& $\kappa_{2z}=\frac{1}{2}
\frac{\dee}{\dee t}
\alpha^2_{01111}$ 
& $\alpha^2_{01111}=	
\left\langle z^2\right\rangle$ \\
\cline{2-3} 			

& $\kappa_{3z}=
\frac{1}{6}\frac{\dee}{\dee t}
\alpha^3_{01111}$ 
& $\alpha^3_{01111}=3\left\langle z \right\rangle
\left\langle z^2 \right\rangle 
-\left\langle z^3 \right\rangle$ \\
\cline{2-3}

& $\kappa_{4z}=\frac{1}{24}
\frac{d}{dt}\alpha^4_{01111}$ 
& $\alpha^4_{01111}=
\left\langle z^4\right\rangle
-3\left\langle z^2\right\rangle^2
-4\langle z\rangle 
\left\langle z^3\right\rangle
+12\langle z\rangle^2
\left\langle z^2\right\rangle$ \\
\cline{2-3} 					

& $\kappa_{5z}=\frac{1}{120}
\alpha^5_{01111}$ 
& $\alpha^5_{01111}=
-\left\langle z^5 \right\rangle
+10 \left\langle z^3\right\rangle
\left\langle z^2 \right\rangle
-30\left\langle z \right\rangle
\left\langle z^2 \right\rangle^2
+5\left\langle z\right\rangle
\left\langle z^4 \right\rangle$\\
& &
$-20\left\langle  z
\right\rangle^2
\left\langle z^3 \right\rangle
+60\left\langle z
\right\rangle^3
\left\langle z^2 \right\rangle$ \\
\cline{2-3} 						
\hline	

\multirow{5}{*}{$\frac{\partial{F}}{\partial{t}}=
	\kappa_{1z}\frac{\partial{F}}
	{\partial{z}}
	+\kappa_{2z}\frac{\partial^2{F}}{\partial{z^2}}
	+\kappa_{3z}\frac{\partial^3{F}}{\partial{z^3}}
	+\kappa_{5z}\frac{\partial^5{F}}{\partial{z^5}}$} 

& $\kappa_{1z}=
\frac{\dee}{\dee t}
\alpha^1_{11101}$ 
& $\alpha^1_{11101}=-\langle z\rangle$ \\
\cline{2-3} 			

& $\kappa_{2z}=
\frac{1}{2}\frac{\dee}{\dee t}
\alpha^2_{11101}$ 
& $\alpha^2_{11101}=\left\langle z^2 \right\rangle
-\langle z\rangle^2$ \\
\cline{2-3}

& $\kappa_{3z}=\frac{1}{6}
\frac{\dee}{\dee t}\alpha^3_{11101}$ 
& $\alpha^3_{11101}=
3\left\langle z \right\rangle\left\langle z^2 \right\rangle
-2\left\langle z \right\rangle^3
-\left\langle z^3 \right\rangle$ \\
\cline{2-3} 					

& $\kappa_{5z}
=\frac{1}{120}\frac{\dee}{\dee t}
\alpha^5_{11101}$ 
& $\alpha^5_{11101}=
-\left\langle z^5 \right\rangle
+5\left\langle  z^4
\right\rangle\langle z\rangle
-20\left\langle z^3
\right\rangle\langle z\rangle^2
-24\langle z\rangle^5
$ \\
& &
$
+10\left\langle  z^3
\right\rangle
\left\langle z^2\right\rangle+60\left\langle z^2
\right\rangle\langle z\rangle^3
-30\left\langle  z^2
\right\rangle^2
\langle z\rangle$ \\
\cline{2-3} 						
\hline	

\multirow{1}{*}{$\frac{\partial{F}}{\partial{t}}=
	\kappa_{5z}\frac{\partial^5{F}}{\partial{z^5}}$} 

& $\kappa_{5z}
=\frac{1}{120}\frac{\dee}{\dee t}
\alpha^5_{00001}$ 
& $\alpha^5_{00001}=
-\left\langle z^5 \right\rangle$ \\
\cline{2-3} 								
\hline

\end{tabular}        	   
\end{center}    

\end{table}

\end{document}